\pdfoutput=1
\documentclass[aps, pra, floatfix, groupedaddress, longbibliography,twocolumn,nobibnotes]{revtex4-2} 

\usepackage{float}
\usepackage[table]{xcolor}
\usepackage{graphicx}
\usepackage{braket}
\usepackage{mathtools}
\usepackage{siunitx}
\usepackage{hyperref}
\usepackage{cleveref}
\usepackage{bibunits}
\defaultbibliography{ECD}
\defaultbibliographystyle{plain}
\crefname{table}{Table}{Tables}
\Crefname{table}{Table}{Tables}
\crefname{figure}{Fig.}{Figures}
\Crefname{figure}{Fig.}{Figures}
\crefname{section}{section}{sections}
\Crefname{section}{Section}{Sections}
\crefname{subsection}{section}{Sections}
\Crefname{subsection}{Section}{Sections}
\usepackage[caption=false]{subfig}

\newcommand{\phantomsubfloat}[1]{
	{
		\captionsetup[subfigure]{labelformat=empty}
		\subfloat[][]{#1}
	}%
}
\usepackage{amssymb}
\usepackage{bm}
\DeclareMathOperator*{\argmax}{argmax} 

\makeatletter
\def\maketitle{
	\@author@finish
	\title@column\titleblock@produce
	\suppressfloats[t]}
\makeatother 

\usepackage{xpatch}

\makeatletter
\patchcmd{\@ssect@ltx}
{\addcontentsline{toc}{#1}{\protect\numberline{}#8}}
{}
{}
{}
\makeatother

\begin{document}
\title{Fast Universal Control of an Oscillator with Weak Dispersive Coupling to a Qubit}

\author{Alec Eickbusch}
\email{alec.eickbusch@yale.edu}
\affiliation{Departments of Applied Physics and Physics, Yale University, New Haven, CT 06520, USA}
\affiliation{Yale Quantum Institute, Yale University, New Haven, CT, USA}

\author{Volodymyr Sivak}
\affiliation{Departments of Applied Physics and Physics, Yale University, New Haven, CT 06520, USA}
\affiliation{Yale Quantum Institute, Yale University, New Haven, CT, USA}

\author{Andy Z. Ding}
\affiliation{Departments of Applied Physics and Physics, Yale University, New Haven, CT 06520, USA}
\affiliation{Yale Quantum Institute, Yale University, New Haven, CT, USA}

\author{Salvatore S. Elder}
\affiliation{Departments of Applied Physics and Physics, Yale University, New Haven, CT 06520, USA}
\affiliation{Yale Quantum Institute, Yale University, New Haven, CT, USA}

\author{Shantanu R. Jha}
\thanks{Current address: Department of Electrical Engineering and Computer Science,	Massachusetts Institute of Technology, Cambridge, MA 02139, USA}
\affiliation{Departments of Applied Physics and Physics, Yale University, New Haven, CT 06520, USA}
\affiliation{Yale Quantum Institute, Yale University, New Haven, CT, USA}

\author{Jayameenakshi Venkatraman}
\affiliation{Departments of Applied Physics and Physics, Yale University, New Haven, CT 06520, USA}
\affiliation{Yale Quantum Institute, Yale University, New Haven, CT, USA}

\author{Baptiste Royer}
\affiliation{Departments of Applied Physics and Physics, Yale University, New Haven, CT 06520, USA}
\affiliation{Yale Quantum Institute, Yale University, New Haven, CT, USA}

\author{S. M. Girvin}
\affiliation{Departments of Applied Physics and Physics, Yale University, New Haven, CT 06520, USA}
\affiliation{Yale Quantum Institute, Yale University, New Haven, CT, USA}

\author{Robert J. Schoelkopf}
\affiliation{Departments of Applied Physics and Physics, Yale University, New Haven, CT 06520, USA}
\affiliation{Yale Quantum Institute, Yale University, New Haven, CT, USA}

\author{Michel H.  Devoret}
\email{michel.devoret@yale.edu}
\affiliation{Departments of Applied Physics and Physics, Yale University, New Haven, CT 06520, USA}
\affiliation{Yale Quantum Institute, Yale University, New Haven, CT, USA}

\date{\today}

\maketitle
\begin{bibunit}[naturemag]
\textbf{A controlled evolution generated by nonlinear interactions is required to perform full manipulation of a quantum system, and such control is only coherent when the rate of nonlinearity is large compared to the rate of decoherence \cite{zurekDecoherenceEinselectionQuantum2003}. As a result, engineered quantum systems typically rely on a bare nonlinearity much stronger than all decoherence rates, and this hierarchy is usually assumed to be necessary. In this work, we challenge this assumption by demonstrating the universal control of a quantum system where the relevant rate of bare nonlinear interaction is comparable to the fastest rate of decoherence. We do this by introducing a novel noise-resilient protocol for the universal quantum control of a nearly-harmonic oscillator that takes advantage of an in-situ enhanced nonlinearity instead of harnessing a bare nonlinearity. Our experiment consists of a high quality-factor microwave cavity with weak-dispersive coupling to a much lower quality superconducting qubit \cite{blaisCircuitQuantumElectrodynamics2021}. By using strong drives to temporarily excite the oscillator, we realize an amplified three-wave-mixing interaction, achieving typical operation speeds over an order of magnitude faster than expected from the bare dispersive coupling  \cite{maQuantumControlBosonic2021b}. Our demonstrations include preparation of a single-photon state with $\bm{98\pm 1(\%)}$ fidelity and preparation of squeezed vacuum with a squeezing level of 11.1 dB, the largest intracavity squeezing reported in the microwave regime. Finally, we also demonstrate fast measurement-free preparation of logical states for the binomial \cite{michaelNewClassQuantum2016} and Gottesman-Kitaev-Preskill (GKP) \cite{gottesmanEncodingQubitOscillator2001} quantum error-correcting codes.}

Many quantum physics experiments strive to operate in a regime where the relevant rate of undriven nonlinear interaction is orders of magnitude larger than the fastest rate of decoherence. Such a hierarchy is at the heart of many different types of engineered quantum systems, including prototypical realizations of cavity quantum electrodynamics with Rydberg atoms \cite{harocheNobelLectureControlling2013}, nonlinear quantum electromechanics \cite{chuQuantumAcousticsSuperconducting2017a}, hybrid superconductor-semiconductor systems \cite{burkardSuperconductorSemiconductorHybridcircuit2020}, and circuit quantum electrodynamics \cite{blaisCircuitQuantumElectrodynamics2021}. With the prevalence of such a hierarchy, it is natural to question if a large native nonlinear interaction strength relative to decoherence is required for high-fidelity operations. This question is especially relevant in cases where engineering a large nonlinearity can be difficult or induce detrimental effects such as enhanced loss mechanisms.

Here, we show the counterintuitive result that a large bare nonlinear interaction strength is not required for coherent control, and we demonstrate this idea for the universal control of a nearly-harmonic oscillator. Such quantum oscillators can be realized as electromagnetic \cite{reagorQuantumMemoryMillisecond2016,romanenkoThreeDimensionalSuperconductingResonators2020} or mechanical \cite{chuQuantumAcousticsSuperconducting2017a, satzingerQuantumControlSurface2018, arrangoiz-arriolaResolvingEnergyLevels2019} bosonic modes, and these are promising platforms for several emerging quantum information applications. High-fidelity control of an oscillator is typically achieved by coupling it to an ancillary qubit in the dispersive regime described by the Hamiltonian $H/\hbar = \chi a^\dagger a \sigma_z/2$ where $a$ is the annihilation operator of oscillator, $\sigma_z$ is the Pauli-Z operator of the ancillary qubit, and $[a, a^\dagger] = 1$. In our proof-of-principle demonstration, the oscillator is realized as the lowest-energy mode of a superconducting microwave cavity, and the ancillary qubit is realized by the lowest two energy levels of a transmon.

State-of-the-art methods for universal control in the dispersive regime include the qubit cavity mapping protocol (qcMAP) \cite{leghtasDeterministicProtocolMapping2013}, the Selective Number-dependent Arbitrary Phase (SNAP) and displacement gate set  \cite{krastanovUniversalControlOscillator2015a,heeresCavityStateManipulation2015,foselEfficientCavityControl2020,kudraRobustPreparationWignernegative2021}, measurement-based methods for oscillator state preparation \cite{wangConvertingQuasiclassicalStates2017}, or model-based pulse shaping such as GRadient Ascent Pulse Engineering
(GRAPE)  \cite{khanejaOptimalControlCoupled2005a,heeresImplementingUniversalGate2017,reinholdControllingErrorCorrectableBosonic2019}. All such methods can perform relevant operations in time comparable to $2\pi/\chi$, so the bare dispersive shift is usually engineered to be orders of magnitude larger than the fastest decoherence rate in the system, $\chi/2\pi \gg \max(\Gamma_2, \Gamma_1, \kappa)$, where $\Gamma_2 = 1/T_2$, $\Gamma_1 = 1/T_1$ are the qubit decoherence and relaxation rates and $\kappa$ is the oscillator relaxation rate  \cite{blaisCircuitQuantumElectrodynamics2021,maQuantumControlBosonic2021b}.

As seen by the $2\pi/\chi$ speed limit, all previous methods of universal oscillator control in the dispersive regime rely on harnessing evolution limited by the bare rate of nonlinear interaction $\chi$. However, such approaches are insufficient in the weak-dispersive regime, $\chi/2\pi \lesssim \max(\Gamma_2, \Gamma_1, \kappa)$, since decoherence would lead to low control fidelity. To overcome this challenge and realize coherent control in the weak-dispersive regime, we use resonant microwave drives to induce phase-space displacements of the oscillator far from the origin relative to zero-point fluctuations. With this, the weak four-wave-mixing interaction is transformed into an effective three-wave-mixing interaction between the oscillator relative to its displaced center-of-mass and the qubit. A similar scheme is used for enhancing mixing processes in quantum parametric amplification \cite{royIntroductionParametricAmplification2016} and optomechanical coupling \cite{aspelmeyerCavityOptomechanics2014}, however in these applications the resulting lower-order interactions around the center-of-mass are linear and thus not universal. In addition, some circuit-QED experiments have harnessed driven four-wave-mixing to generate enhanced three-wave-mixing interactions between an oscillator and qubit at rates faster than the native dispersive shift \cite{murchCavityAssistedQuantumBath2012,pechalMicrowaveControlledGenerationShaped2014,eddinsStroboscopicQubitMeasurement2018,rosenblumCNOTGateMultiphoton2018b,rosenblumFaulttolerantDetectionQuantum2018b,touzardGatedConditionalDisplacement2019, reinholdControllingErrorCorrectableBosonic2019,campagne-ibarcqQuantumErrorCorrection2020,elderHighFidelityMeasurementQubits2020,vrajitoareaQuantumControlOscillator2020a}, however high-fidelity universal control faster than $2\pi/\chi$ has not previously been demonstrated.

\begin{figure}[!ht]
	\vspace{-2\baselineskip}
	\phantomsubfloat{\label{fig:ECD control a}}
	\phantomsubfloat{\label{fig:ECD control b}}
	\phantomsubfloat{\label{fig:ECD control c}}
	\phantomsubfloat{\label{fig:ECD control d}}
	\includegraphics{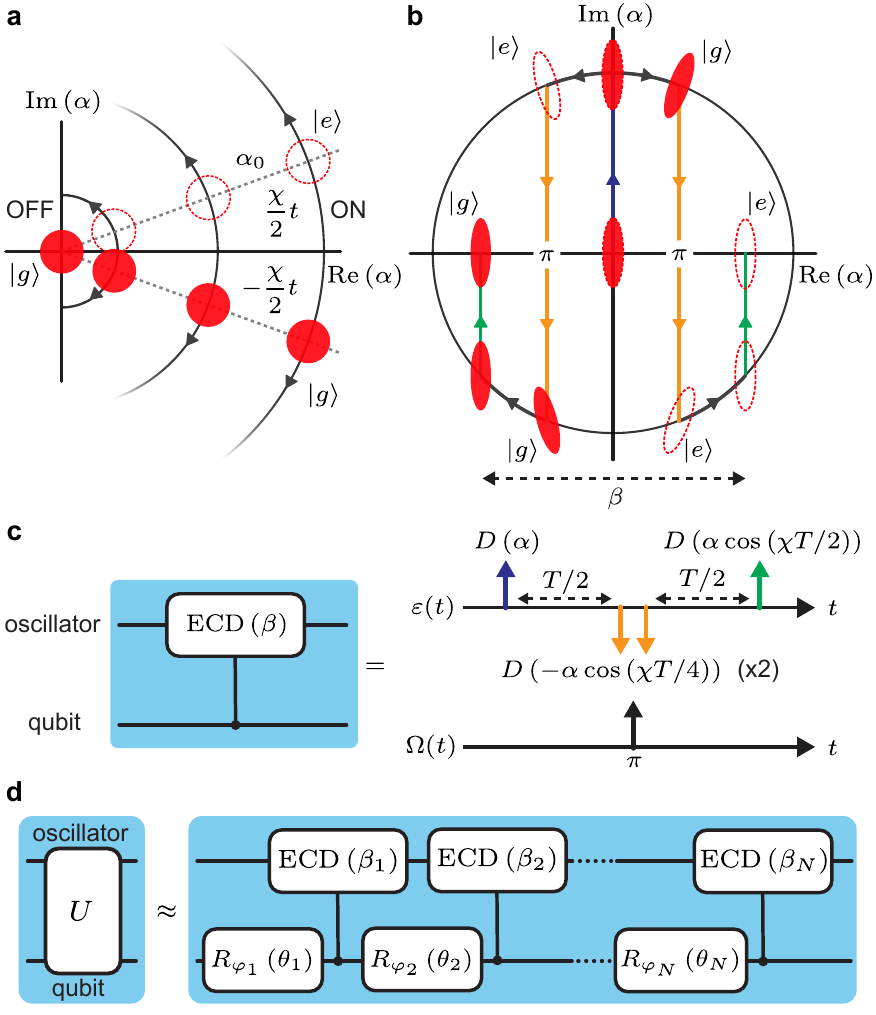}
	\caption{\label{fig:ECD control} \textbf{Echoed Conditional Displacement Control.} \textbf{(a)} Under a conditional rotation, phase-space displacements act as lever arms, generating a large separation conditioned on qubit states $\ket{g}$ and $\ket{e}$. Here, three displacements are shown acting on vacuum, and with a larger $\alpha_0$ there is a larger state separation after a time $t$.  \textbf{(b)} The ECD gate phase-space trajectory in the limit of instantaneous displacements acting on a squeezed state for illustration. \textbf{(c)} Oscillator drive ($\varepsilon(t)$) and qubit drive ($\Omega(t)$) for the ideal ECD gate of duration $T$ resulting in a final state separation of $\beta = 2 i\alpha \sin(\chi T/2)$ where $|\alpha| = \alpha_0$. \textbf{(d)} Any unitary can be approximated by a sequence of $N$ single qubit rotations and ECD gates with built-in dynamical decoupling generated through the symmetric construction of the full sequence.}
\end{figure}

The enhancement relies on a phase-space displacement acting as a lever arm under the dynamics of the dispersive interaction as shown in \cref{fig:ECD control a}. To analyze this effect, the dispersive Hamiltonian $H$ can be transformed into a displaced frame, giving

\begin{equation}
	\label{eq:displaced H}
	\tilde{H}/\hbar = \chi a^\dagger a \frac{\sigma_z}{2} + \chi(\alpha(t) a^\dagger + \alpha^*(t) a) \frac{\sigma_z}{2} + \chi |\alpha(t)|^2 \frac{\sigma_z}{2},
\end{equation}

where $\partial_t \alpha(t) = -i\varepsilon(t)- \left(\kappa/2\right)\alpha(t)$ is the classical response to a resonant drive, $H_d/\hbar = \varepsilon^*(t) a + \varepsilon(t) a^\dagger$. With a large displacement $\alpha_0 = \text{max}|\alpha(t)|$, the second term in $\tilde{H}$ dominates, and the effective interaction between the oscillator and qubit becomes a qubit-state-dependent force with maximum effective interaction strength $g_\text{eff} = \chi \alpha_0$ \cite{murchCavityAssistedQuantumBath2012,eddinsStroboscopicQubitMeasurement2018,campagne-ibarcqQuantumErrorCorrection2020}. The other terms in $\tilde{H}$ also contribute to the dynamics, and their effect must be accounted for or canceled with a suitable qubit echo sequence such as the gate construction introduced below.

For a transmon with anharmonicity $K$, the critical oscillator photon number limits the enhanced interaction to $g_\text{eff}^\text{max} \approx \sqrt{\chi K/6}$, setting a maximum speed limit of universal control using this approach (see Methods). In addition, it should be noted that large displacements populate highly excited states of the oscillator which can enhance decoherence mechanisms. For superconducting cavities, photon loss is the dominant error channel \cite{reagorQuantumMemoryMillisecond2016}, and under a coherent displacement there is no enhanced decoherence due to relaxation provided the deterministic re-centering force at a rate $\kappa/2$ is included when calculating the drive $\varepsilon(t)$ needed for a desired oscillator trajectory $\alpha(t)$. However, oscillator dephasing at a rate $\kappa_\phi$ does cause enhanced decoherence under a displacement, resulting in diffusion-like terms at an effective rate $2|\alpha(t)|^2\kappa_\phi$ (see Methods). Oscillators such as superconducting cavities can have dephasing rates much weaker than their relaxation rates limiting this effect \cite{reagorQuantumMemoryMillisecond2016, rosenblumFaulttolerantDetectionQuantum2018b, campagne-ibarcqQuantumErrorCorrection2020}, however it reveals a trade-off between faster control with large displacements and enhanced loss from oscillator dephasing.

With this in mind, the enhanced three-wave-mixing interaction and a qubit $\pi$ pulse can be used to engineer an entangling gate dubbed the echoed conditional displacement (ECD) gate, defined as $\text{ECD}(\beta) = D(\beta/2)\ket{e}\bra{g} + D(-\beta/2)\ket{g}\bra{e}$, where $D(\alpha) = e^{\alpha a^\dagger - \alpha^* a}$ is the displacement operator. A version of the gate was first implemented as a tool to realize error correction of GKP states \cite{campagne-ibarcqQuantumErrorCorrection2020}, and with an intermediate oscillator displacement of length $\alpha_0$, the $\text{ECD}(\beta)$ gate occurs in a time approximately $|\beta|/\left(\chi \alpha_0\right)$ through the trajectory shown in \cref{fig:ECD control b} and ideal drive sequence in \cref{fig:ECD control c}. The ECD gate cancels the dynamics from the dispersive and Stark shift terms in $\tilde{H}$ up to a qubit phase because of its symmetric construction and qubit echo (see Supplementary Information section 4).

\begin{figure*}[!ht]
	\vspace{-2\baselineskip}
	\phantomsubfloat{\label{fig:fock state creation a}}
	\phantomsubfloat{\label{fig:fock state creation b}}
	\phantomsubfloat{\label{fig:fock state creation c}}
	\phantomsubfloat{\label{fig:fock state creation d}}
	\phantomsubfloat{\label{fig:fock state creation e}}
	\includegraphics{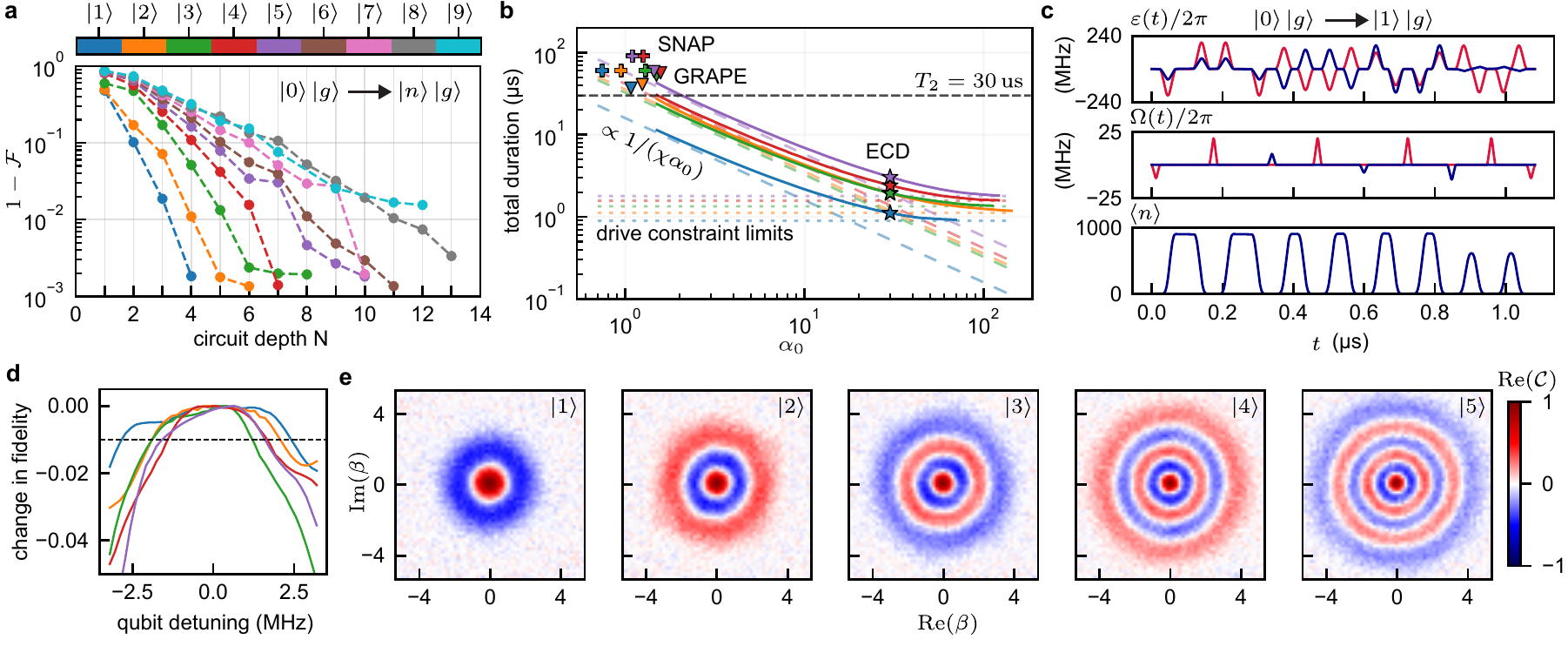}
	\caption{\label{fig:fock state creation} \textbf{Fock state preparation.} \textbf{(a)} Best state transfer infidelity found when optimizing ECD circuit parameters to prepare oscillator Fock state $\ket{n}$ from vacuum as a function of circuit depth $N$. Here, $\mathcal{F} = \left|\bra{g}\bra{n} U_\text{ECD} \ket{0}\ket{g}\right|^2$. \textbf{(b)} Total pulse sequence duration using the protocols from (a) with minimum $N$ such that $\mathcal{F} \geq 0.99$ (solid lines). Color code is same as in part (a). Colored long-dashed lines are the instantaneous displacement scaling $T_\text{total} = \left(\chi \alpha_0\right)^{-1} \sum_{i=1}^N |\beta_i|$. Colored dotted lines are the drive constraint limits $T_\text{total} = 2N t_q + 4N t_D$. We use $\alpha_0 = 30$ in our experiment as indicated by the stars. Also included are durations for independently optimized GRAPE (triangles) and SNAP (crosses) protocols using our system parameters where the x-coordinate indicates the simulated largest displacement used ($\max|\braket{a}|$). \textbf{(c)} Cavity drive $\varepsilon(t)$ (upper panel) and transmon drive $\Omega(t)$ (middle panel) for preparation of Fock state $\ket{1}$ (real and imaginary parts shown in red and blue) and simulated average photon number during the sequence (lower panel). \textbf{(d)} Simulated drop in fidelity of Fock preparation sequences with a qubit detuning $H/\hbar = \delta \sigma_z/2$. Dashed line indicates a drop of $1$\%. \textbf{(e)} Measured real part of the characteristic functions for the first five excited Fock states in the cavity. Preparation fidelities are given in \cref{table1}.}
\end{figure*}

\sloppy To build a full gate set, we combine the $\text{ECD}(\beta)$ gate with unselective qubit rotations, $R_\varphi(\theta) = \exp\left[-i(\theta/2)\left(\sigma_x \cos\varphi + \sigma_y \sin\varphi\right)\right]$. The rotation pulse bandwidth must be sufficiently large compared to $\braket{a^\dagger a}\chi$ so the oscillator state does not entangle with the qubit during the rotations, additionally requiring $K\gg\chi$ to avoid populating higher excited states of the transmon. Together the set $\left\{\text{ECD}(\beta),R_\varphi(\theta) \right\}$ is universal for control of the oscillator and qubit (see Methods). Any desired unitary on the joint oscillator and qubit Hilbert space can be approximated by the decomposition shown in \cref{fig:ECD control d}, with $4$ real-valued parameters per step, and a fidelity that depends on the circuit depth $N$. The full sequence has total duration $T_\text{total} = \left(\chi \alpha_0\right)^{-1} \sum_{i=1}^N |\beta_i|$ in the instantaneous displacement and qubit pulse limit, hence a large $\alpha_0 \gg 1$ can enhance the overall speed of a target unitary. Every ECD control sequence has intrinsic dynamical decoupling of low-frequency noise coupled to $\sigma_z$ because of its designed symmetric structure, motivating the choice of this gate set.

To realize a desired unitary or state-transfer with the ECD gate-set, we use a two-step optimization approach. In the first step, we find the circuit parameters $\{\vec{\beta}, \vec{\varphi}, \vec{\theta}\}$ that maximize the fidelity and minimize the circuit depth $N$. Here, we use an efficient gradient-based parameter optimization using automatic differentiation implemented on a graphics processing unit (see Supplementary Information section 7). We note that the circuit parameters could also be applied to realize universal control of the motional state of a trapped ion, where a conditional displacement gate can also be realized \cite{haljanSpinDependentForcesTrapped2005}.

As an example of this first step of optimization, we focus on the preparation of Fock states in the oscillator, $\ket{0}\ket{g} \rightarrow \ket{n}\ket{g}$. These are not simple superpositions of displaced coherent states, so it is not obvious that they can be easily prepared using conditional displacements starting from vacuum. Optimization results for preparation of Fock states $\ket{1}$ through $\ket{9}$ are shown in \cref{fig:fock state creation a}. The circuit depth required increases with photon number, with 10 or fewer ECD gates needed to reach a state transfer fidelity $\mathcal{F} > 0.99$ for the first seven Fock states. This example shows that ECD control can be an efficient circuit parameterization, with only $4N$ total parameters per sequence, a circuit depth comparable to the SNAP protocol \cite{krastanovUniversalControlOscillator2015a, foselEfficientCavityControl2020,kudraRobustPreparationWignernegative2021}, and over an order of magnitude fewer parameters than time-domain GRAPE as used in state-of-the-art bosonic experiments \cite{maQuantumControlBosonic2021b}.

In the second optimization step, the cavity drive $\varepsilon(t)$ and qubit drive $H_q/\hbar = \Omega^*(t)\sigma^- + \Omega(t) \sigma^+$ are compiled from a set of ECD circuit parameters found in the first step. This optimization is done with realistic constraints to realize the ECD sequence in the shortest time given bandwidth and amplitude limits (see Supplementary Information section 3 and 4). In our experiment, we use a 3D aluminum superconducting cavity (frequency $\SI{5.26}{GHz}$, relaxation time $T_{1,c} = \SI{436}{us}$) coupled to a transmon qubit (frequency $\SI{6.65}{GHz}$, relaxation time $T_{1,q} \approx \SI{50}{us}$) with a dispersive shift $\chi/2\pi = \SI{33}{kHz}$. Given this bare nonlinearity, the resulting sequence duration for preparation of Fock states $\ket{1}$ through $\ket{5}$ as a function of the displacement used during the ECD gates $\alpha_0$ is shown in \cref{fig:fock state creation b}. At intermediate values of $\alpha_0$ the sequence duration follows the instantaneous-displacement limit $T_\text{total} \propto \left(\chi \alpha_0\right)^{-1}$. As $\alpha_0$ increases, the amplitude and bandwidth constraints result in sequences limited by the total duration of the constituent pulses, $T_\text{total} = 2N t_q + 4N t_D$, where the duration of qubit rotation pulse and oscillator displacement pulses used in our experiment are $t_q = \SI{24}{ns}$ and $t_D = \SI{44}{ns}$ (see Supplementary Information section 4). From our transmon anharmonicity of $K/2\pi = \SI{181}{MHz}$, the maximum conditional displacement rate is $g_\text{eff}^\text{max}/2\pi \approx \SI{1}{MHz}$. In our experimental demonstrations, we use $\alpha_0 = 30$ (as shown by the stars in \cref{fig:fock state creation b}) and operate close to this bound. Finally, the shortest lifetime in our experiment is the transmon Ramsey coherence time $T_2 \approx \SI{30}{us}$ realizing $\chi/2\pi \lesssim \Gamma_2 \ll g_\text{eff}/2\pi$ and allowing high-fidelity control in a regime where the bare nonlinearity is comparable to the fastest decoherence rate.

In \cref{fig:fock state creation b}, the duration of ECD pulse sequences are also compared to independently optimized GRAPE and SNAP sequences for Fock state preparation with our system parameters (see Supplementary Information section 9). Here, ECD sequences have over an order-of-magnitude enhancement in gate speed. For example, single-photon state preparation in the oscillator is realized about $30$ times faster than $2\pi / \chi$, with compiled drives and simulated intracavity average photon number shown in \cref{fig:fock state creation c}, demonstrating the ability to utilize the oscillator's vast Hilbert space to enhance gate speed with a displaced-field of $\text{max}|\alpha|^2 = 900$ photons during the gates. Finally, as a simple analysis to indicate the insensitivity to low-frequency noise coupled to $\sigma_z$, \cref{fig:fock state creation d} shows the result of simulating the ECD Fock state preparation pulse sequences with an additional qubit detuning $H/\hbar =\delta \sigma_z/2$, showing resilience at the level of $1-\mathcal{F} \sim 0.01$ to static detuning on the order of $\delta/\hbar \sim \SI{1}{MHz}$. 

To characterize the performance of these protocols applied in experiment, we measure the complex-valued characteristic function $\mathcal{C}(\beta) = \text{Tr}\left(D(\beta)\rho \right)$ by using an ECD gate to perform phase estimation of the displacement operator $D(\beta)$ conditioned on a disentangling measurement (see Supplementary Information section 5) \cite{campagne-ibarcqQuantumErrorCorrection2020, fluhmannDirectCharacteristicFunctionTomography2020, fluhmannEncodingQubitTrappedion2019}. Importantly, tomography can also be performed in a time much faster than $2\pi/\chi$ using large displacements (we note that direct Wigner tomography using typical circuit-QED parity measurements would be impractical, taking a time $\pi/\chi \approx \SI{15}{us}$.) The real parts of the measured characteristic functions for Fock states $\ket{1}$ through $\ket{5}$ are shown in \cref{fig:fock state creation e}. From the real and imaginary parts (not shown) of the characteristic functions we reconstruct the density matrices using maximum likelihood estimation leading to the results summarized in \cref{table1} and reaching a best fidelity of  $\mathcal{F}_\text{exp} = 0.98 \pm 0.01$ for Fock state $\ket{1}$.

\begin{table}[!ht]
	\begin{tabular}{c|cccc}
		State& $\mathcal{F}_\text{exp}$ (\%) & $\mathcal{F}_{\text{sim}}$ (\%) &  $\mathcal{F}_{\text{sim}}^{\kappa_\phi}$ (\%)\\
		\hline 
		$\ket{1}$ & 98     & 99 &98 \\
		$\ket{2}$ &  92     & 97&94 \\
		$\ket{3}$ & 88      & 97 &93 \\
		$\ket{4}$ & 87     &97 &92 \\
		$\ket{5}$ & 82      & 94&83 \\
		\hline
		$\ket{+Z}_\text{bin}$ &  92   & 98 &95      \\
		$\ket{+X}_\text{bin}$  &89     &  97 &94  \\
		$\ket{+Y}_\text{bin}$   & 91    &  97  &93       \\
		\hline
		$\ket{+Z}_\text{GKP}$        &   85   & 93& 85  \\
		$\ket{+Y}_\text{GKP}$        &  83  & 92& 87  \\
		$\ket{-Z}_\text{GKP}$        &   80 & 93& 85  \\
	\end{tabular}
	\caption{\label{table1} \textbf{Measured and simulated state preparation fidelities.} $\mathcal{F}_\text{exp}$ is the measured fidelity found from density matrix reconstruction, $\mathcal{F}_{\text{sim}}$ is the simulated fidelity including all independently measured decoherence rates, and $\mathcal{F}_{\text{sim}}^{\kappa_\phi}$ is the simulated fidelity including additional cavity dephasing at a rate \textbf{ $\kappa_\phi = \left(\SI{150}{ms}\right)^{-1}$}. Fidelity is defined as $\mathcal{F} = \bra{\psi_t}\rho_g\ket{\psi_t}$ where $\rho_g$ is the oscillator state after projecting the qubit in $\ket{g}$ and $\ket{\psi_t}$ is the oscillator target state. We estimate the quoted fidelities are accurate within $1\%$ using bootstrapping. The average probability of measuring $\ket{g}$ after the state preparation sequences are $\left\{0.96,0.93,0.96,0.92\right\}$ for the Fock, squeezed, binomial, and GKP states respectively.}
\end{table}

In \cref{table1} the fidelity of ECD sequences measured in experiment are compared to master equation simulations $\mathcal{F}_\text{sim}$ including all independently measured decoherence mechanisms (see Supplementary Information section 6). Out of the measured decoherence sources, qubit relaxation during the ECD gates is the largest contribution to the simulated infidelity, as this error channel can lead to unknown displacements of the oscillator during the sequence. However, for most demonstrations, $\mathcal{F}_\text{sim}$ overestimates the measured fidelity. To investigate a likely additional decoherence mechanism, we also simulate the protocols including oscillator dephasing at a rate $\kappa_\phi = (\SI{150}{ms})^{-1}$, resulting in fidelity of sequences given by $\mathcal{F}_\text{sim}^{\kappa_\phi}$ in \cref{table1}, included here to demonstrate the sensitivity to a small oscillator dephasing rate that is enhanced with the large displacements. Finally, additional infidelity in the experiment can arise from unknown microwave transfer functions and other forms of model bias. However, the use of a few-parameter gate-set allows for the implementation of model-free closed-loop optimization strategies such as reinforcement learning \cite{sivakModelFreeQuantumControl2021,baumExperimentalDeepReinforcement2021b} which could be used in future investigations to combat these effects. 

\begin{figure*}[!ht]
	\vspace{-2\baselineskip}
	\phantomsubfloat{\label{fig:bosonic code prep a}}
	\phantomsubfloat{\label{fig:bosonic code prep b}}
	\phantomsubfloat{\label{fig:bosonic code prep c}}
	\phantomsubfloat{\label{fig:bosonic code prep d}}
	\includegraphics{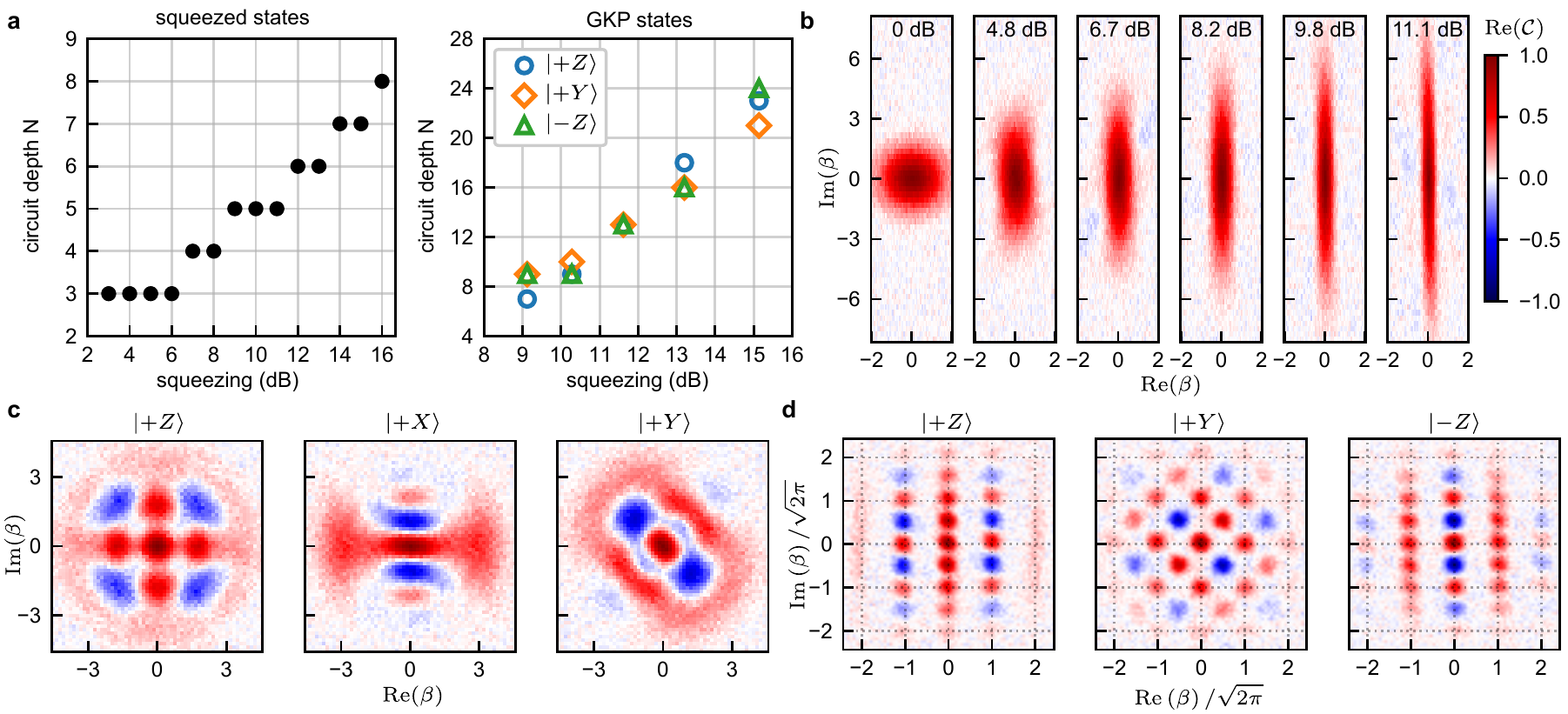}
	\caption{\label{fig:bosonic code prep} \textbf{Preparation of squeezed states and bosonic code words.} \textbf{(a)} Numerically optimized minimum circuit depth to reach state transfer fidelity $\mathcal{F} \geq 0.99$ for preparation of squeezed states (left panel) and $\mathcal{F} \geq 0.98$ for preparation of logical GKP states (right panel) \textbf{(b)} Real part of measured characteristic functions for vacuum and squeezed states, with achieved squeezing indicated. \textbf{(c) and (d)} Measured characteristic functions for logical state preparation of binomial (c) and GKP (d) code words. For the binomial code, $\ket{-Z_L} = \ket{2}$ is included in \cref{fig:fock state creation e}. For both codes, all other cardinal points on the Bloch sphere can be obtained by phase space rotations. Fidelities for the binomial and GKP code are given in \cref{table1} with additional analysis in Supplementary Information section 5.}
\end{figure*}

As a second demonstration, we prepare squeezed vacuum states $\ket{\zeta} = \exp\left(\frac{1}{2}(\zeta^* a^2 - \zeta a^{\dagger 2})\right) \ket{0}$ with a squeezing level in dB defined as $20 \log_{10}\left(e^{\left|\zeta\right|}\right)$. Highly squeezed states of an oscillator can allow sensing beyond the standard quantum limit, as was recently used to enhance the search for dark matter axions \cite{backesQuantumEnhancedSearch2021}. However, the presence of a large oscillator self-Kerr degrades the quality by distorting the squeezed state and increasing the variance of the squeezed quadrature \cite{dassonnevilleDissipativeStabilizationSqueezing2021}. In our experiment, the small inherited oscillator self-Kerr of $\approx\SI{1}{Hz}$, over three orders of magnitude smaller than is typically used \cite{heeresImplementingUniversalGate2017}, minimizes the state distortion during preparation and idling periods.

In the left panel of \cref{fig:bosonic code prep a} we show ECD optimization results for squeezed state preparation starting from vacuum. Plotted is the the minimum circuit depth needed to reach a fidelity $\mathcal{F} \geq 99\%$. A related method for squeezed state creation was introduced in \cite{hastrupUnconditionalPreparationSqueezed2021a}, and the protocols here have fewer conditional displacements, demonstrating the ability of our optimization method to find novel control circuits. In our experiment, we apply the optimized squeezed state preparation ECD sequences for target squeezing levels of $\left\{6,8,10,12,14\right\}$ dB using $\alpha_0 = 30$. The measured characteristic functions are shown in \cref{fig:bosonic code prep b}, with achieved squeezing levels of $\left\{4.8,6.7,8.2,9.8,11.1\right\}$ dB calculated from a fit to the reconstructed probability distribution along the squeezed quadrature (see Supplementary Information section 5). The reconstructed states show some non-Gaussian features as decoherence during the pulse causes distortion, similar to the Fock preparation case, leading to a lower effective squeezing. To the best of our knowledge, the measured squeezing of $11.1$ dB demonstrated here is larger than any intraresonator squeezing demonstration in the microwave regime to date, with other demonstrations achieving steady-state intracavity squeezing at the level of $8.2$ dB \cite{dassonnevilleDissipativeStabilizationSqueezing2021} and a postselected state-preparation method demonstrating $5.7$ dB \cite{wangConvertingQuasiclassicalStates2017}. The echoed conditional displacements realized here could also be used to sense small displacements of a squeezed state using phase estimation \cite{duivenvoordenSinglemodeDisplacementSensor2017}.

Finally, to apply the protocol to generate more complex non-Gaussian states, we implement logical state preparation for two different quantum error correcting bosonic codes, the binomial code \cite{michaelNewClassQuantum2016} and the square GKP code \cite{gottesmanEncodingQubitOscillator2001}, as the fast initialization of logical code states is an important resource. For the binomial code, we focus on the smallest code for which the loss of a single photon is correctable, with code words $\ket{+Z}_\text{bin} = \left(\ket{0} + \ket{4}\right)/\sqrt{2}$ and $\ket{-Z}_\text{bin} = \ket{2}$. The GKP code, on the other hand, is defined as the mutual +1 eigenspace of the displacement stabilizers $S_p = D(\sqrt{2\pi})$ and $S_q = D(i\sqrt{2\pi})$ with logical operators given by $X = D(\sqrt{\pi/2})$ and $Z = D(i\sqrt{\pi/2})$. The ideal GKP code has infinite energy, and a finite energy code can be defined by modifying the stabilizers and logical operators using the envelope operator $E_\Delta = \exp{\left\{-\Delta^2 a^\dagger a\right\}}$ under the transformation $O_\Delta = E_\Delta O E_\Delta^{-1}$, leading to code states that are superpositions of squeezed states with squeezing parameter $\zeta = \ln \Delta$ \cite{nohQuantumCapacityBounds2019, royerStabilizationFiniteEnergyGottesmanKitaevPreskill2020}.

For the binomial code, previous experiments have demonstrated logical state preparation using GRAPE, relying on a large bare nonlinearity $\chi/2\pi$ compared to decoherence rates \cite{heeresImplementingUniversalGate2017,axlineOndemandQuantumState2018,huQuantumErrorCorrection2019, gertlerProtectingBosonicQubit2021,burkhartErrorDetectedStateTransfer2021}. With ECD control, optimization results in protocols that prepare all cardinal points of the Bloch sphere to a fidelity $\mathcal{F} = 99\%$ with a circuit depth at most $N = 5$, and the approach is compatible with the low-$\chi$ regime. Applying these circuits in experiment using $\alpha_0 = 30$ results in the measured characteristic functions shown in \cref{fig:bosonic code prep c} with fidelity given in \cref{table1}. The average pulse time for binomial state preparation is $\SI{3.27}{us}$ - about 9 times faster than $2\pi/\chi$. In principle, fast logical operations, measurement, and stabilization of the binomial code could also be performed using ECD control.

To prepare GKP states, the number of conditional displacements required increases with the code squeezing. In the right panel of \cref{fig:bosonic code prep a} we plot the circuit depth required for ECD protocols to prepare $\ket{+Z}_\text{GKP}$, $\ket{+Y}_\text{GKP}$, and $\ket{-Z}_\text{GKP}$ optimized to a state transfer fidelity of $\mathcal{F} = 98\%$ at different squeezing levels. The protocols found here represent a unitary protocol for GKP state creation, as opposed to the non-unitary protocols recently demonstrated in both trapped ions \cite{fluhmannEncodingQubitTrappedion2019} and superconducting circuits \cite{campagne-ibarcqQuantumErrorCorrection2020} that require multiple measurements with feedback or many rounds of dissipative pumping. A related measurement-free GKP state preparation protocol has been proposed \cite{hastrupMeasurementfreePreparationGrid2021} and implemented in trapped-ions \cite{deneeveErrorCorrectionLogical2020}, however it requires an initial squeezed state.

In \cref{fig:bosonic code prep d}, we plot the measured characteristic functions found using these circuits with $\alpha_0=30$ achieving fidelities given in \cref{table1}. For the GKP states, we use a target squeezing level of 10.3 dB and experimentally achieve a squeezing level of 9.1 dB (see Supplementary Information section 5). Our pulse sequences are about 15 times faster than the state preparation method using measurements and feedback demonstrated in \cite{campagne-ibarcqQuantumErrorCorrection2020} with similar experimental parameters. This order-of-magnitude reduction in initialization time can reduce the hardware overhead of error correction protocols requiring GKP resource states, such as teleported error correction \cite{grimsmoQuantumErrorCorrection2021} or the GKP surface code \cite{vuillotQuantumErrorCorrection2019,nohLowOverheadFaultTolerantQuantum2022}.

The experimental demonstrations in this work have focused on oscillator state preparation, however the ECD protocol is universal and can also be extended to performing fast unitary gates. As a demonstration of this, in Supplementary Information section 8 we show numerical optimization of the logical $S = \text{diag}\left(1,e^{i\pi/2}\right)$ and $T = \sqrt{S} = \text{diag}\left(1,e^{i\pi/4}\right)$ gates for a finite-energy GKP code at different squeezings $\Delta$. Remarkably, a circuit depth of only $N=3$ is required to reach a gate fidelity of $\overline{\mathcal{F}} \approx 0.99$ for the $T$ gate and $N=4$ for the $S$ gate at $\Delta = 0.25$, revealing that the ECD gate set is well suited for control over the finite-energy GKP code.

With these proof-of-principle results, we demonstrate the counterintuitive result that high-fidelity universal control can be carried out in a regime where the relevant rate of bare nonlinear evolution is comparable to the fastest decoherence rate. In particular, a large on-off ratio between the rate of control and the bare oscillator-qubit hybridization can be achieved without the need for additional hardware such as a tunable coupler. Importantly, the approach still requires a large ancilla qubit nonlinearity K, reflected by the enhanced interaction speed limit $\propto \sqrt{\chi K}$. 

Although our examples are specific to the oscillator and qubit system, similar displaced-field type control schemes could be designed and performed in other bosonic systems with bare nonlinearity of fourth order or greater, such as the recently proposed scheme to enhance the rate of Fock state preparation in a Kerr oscillator \cite{lingenfelterUnconditionalFockState2021a}. Additionally, using a weak bare nonlinearity has many benefits in the context of quantum information processing - for example, by sufficiently reducing the dispersive coupling $\chi$, oscillator nonlinearity and loss inherited from the qubit can be minimized while retaining controllability, realizing a modular architecture where the qubit and oscillator can be more independently optimized. This is important in applications where these spurious couplings can cause decoherence and distortion of encoded states, especially during idling periods \cite{albertPerformanceStructureSinglemode2018}. Also, the approach could allow for control of oscillators with measured relaxation times on the order of seconds \cite{romanenkoThreeDimensionalSuperconductingResonators2020} without reducing their lifetimes from the coupling to a lossy qubit through the Purcell effect \cite{blaisCircuitQuantumElectrodynamics2021}. Other applications include control in architectures where a large dispersive coupling can be difficult to realize, such as semiconducting qubits coupled to microwave resonators \cite{burkardSuperconductorSemiconductorHybridcircuit2020} and superconducting qubits coupled to acoustic resonators \cite{chuQuantumAcousticsSuperconducting2017a,satzingerQuantumControlSurface2018,arrangoiz-arriolaResolvingEnergyLevels2019}, or as a means to selectively control single bosonic modes when multiple modes are coupled to the same ancillary controller \cite{chakramSeamlessHighMicrowave2021}.

\section*{Methods}

\subsection*{Oscillator relaxation and dephasing in the displaced frame}

With photon loss at a rate $\kappa$, the oscillator's density matrix evolves according to the quantum master equation in Lindblad form
\begin{align}
	\partial_t \rho &= -i \left[H,\, \rho\right] + \kappa \mathcal{D}\left[a\right]\rho, \\
	H &= H_0 + \varepsilon  a^\dagger + h.c.,
\end{align}
were $\mathcal{D}[L] = L\rho L^\dagger - \left(1/2\right)\left\{L^\dagger L, \rho\right\}$, $H_0$ is the oscillator's Hamiltonian, and we have included a time-dependent oscillator drive $\varepsilon(t)$. Here, we take $\hbar = 1$.
Evolution of the density matrix in a time-dependent displaced frame $\tilde{\rho} = D^\dagger \left(\alpha\right) \rho D\left(\alpha\right)$ is given by the equivalent master equation
\begin{align}
	\partial_t \tilde{\rho} &= -i\left[\tilde{H}, \tilde{\rho} \right] + \kappa \mathcal{D}\left[a + \alpha\right]\tilde{\rho} \\
	\tilde{H} &= D^\dagger \left(\alpha\right)H_0 D\left(\alpha\right) + \left( -i \partial_t \alpha + \varepsilon \right)a^\dagger + h.c.
\end{align}
In particular, the displaced frame Lindbladian can be recast as
\begin{equation}
	\kappa \mathcal{D}\left[a + \alpha\right]\tilde{\rho} = \kappa \mathcal{D}\left[a\right]\tilde{\rho}  - i \left[i\frac{\kappa}{2} \left(\alpha^* a - \alpha a^\dagger\right), \tilde{\rho}\right],
\end{equation}
corresponding to photon loss at a rate $\kappa$, and a Hermitian re-centering force at a rate $\frac{\kappa}{2}|\alpha|$. This deterministic force can be lumped into the effective displaced-frame Hamiltonian, giving
\begin{align}
	\partial_t \tilde{\rho} &= -i\left[\tilde{\tilde{H}}, \tilde{\rho} \right] + \kappa \mathcal{D}\left[a\right]\tilde{\rho} \\
	\tilde{\tilde{H}} &= D^\dagger \left(\alpha\right)H_0 D\left(\alpha\right) + \left( -i \partial_t \alpha - i \frac{\kappa}{2} \alpha + \varepsilon \right)a^\dagger + h.c.
\end{align}
Given a desired $\alpha(t)$, $\varepsilon(t)$ can be chosen such that the term in parentheses is zero, satisfying the classical Langevin equation for $\alpha(t)$ given in the main text and counteracting the re-centering force. With this choice of drive, the deterministic evolution is accounted for, and relaxation in the displaced frame is not enhanced compared to relaxation at the origin of phase space. The classical drive equation can also be modified to account for all linear terms in $\tilde{H}$, including those caused by nonlinear terms in $H_0$ (see Supplementary Information section 4).

White-noise oscillator dephasing is given by the master equation $\partial_t \rho = 2\kappa_\phi\mathcal{D}\left[a^\dagger a\right]\rho$. Defining the superoperator $\mathcal{S}\left[X,Y\right]\rho = X \rho Y^\dagger - \left\{Y^\dagger X, \rho \right\}$, oscillator dephasing is transformed in the displaced frame to 
\begin{equation}
	\begin{aligned}
		&\partial_t \tilde{\rho} = 2\kappa_\phi \mathcal{D}[(a^\dagger + \alpha^*)(a + \alpha)]\tilde{\rho} \\
		&=2\kappa_\phi \left\{ \mathcal{D}[a^\dagger a]\tilde{\rho} + |\alpha|^2 \left(\mathcal{D}[a]\tilde{\rho} + \mathcal{D}[a^\dagger]\tilde{\rho}\right)\right. \\
		&+ \alpha^{2}\mathcal{S}\left[a^\dagger, a\right]\tilde{\rho} + \alpha^{*2}\mathcal{S}\left[a, a^\dagger\right]\tilde{\rho} \\
		&+ \alpha \left(\mathcal{S}\left[a^\dagger a, a\right]\tilde{\rho} + \mathcal{S}\left[a^\dagger,  a^\dagger a\right]\tilde{\rho}\right)\\
		&\left.+ \alpha^* \left(\mathcal{S}\left[a^\dagger a, a^\dagger\right]\tilde{\rho} + \mathcal{S}\left[a,  a^\dagger a\right]\tilde{\rho}\right)\right\}.
	\end{aligned}
\end{equation}
In the displaced frame, the noise is dominated by diffusion-like terms at rate $2\kappa_\phi |\alpha|^2$, and unlike the relaxation case, there is no deterministic part that can be counteracted with a simple displacement. However, this master equation is only valid in the Markovian regime, and typically the spectral density of oscillator frequency fluctuations non-white due to effects such as two-level-system defects \cite{niepceStabilitySuperconductingResonators2021}. In the colored noise case, it is possible that part of the enhanced dephasing noise could be echoed away using symmetric pulse constructions \cite{ballSoftwareToolsQuantum2021}.

\subsection*{Universality of ECD Control}
Universal control of the oscillator is the ability to perform arbitrary unitary transformations which are generated by Hamiltonians polynomial in $q =(1/\sqrt{2})(a^\dagger + a)$ and $p = (i/\sqrt{2})(a^\dagger - a)$ \cite{lloydQuantumComputationContinuous1999b, braunsteinQuantumInformationContinuous2005}. Here, we extend this definition to universal control of the oscillator and qubit, which is the ability to perform arbitrary unitary transformations which are generated by linear combinations of Hamiltonians of the form $q^j p^k \sigma_i$ where $j,k$ are non-negative integers and $\sigma_i \in \left\{I, \sigma_x, \sigma_y, \sigma_z\right\}$.

Given a set of generating Hamiltonians $\left\{A,B\right\}$, the two identities
\begin{align}
	e^{-iA \delta t}e^{-iB \delta t}e^{iA \delta t}e^{iB \delta t} &= e^{\left[A,B\right]\delta t^2} + O(\delta t^3), \\
	e^{iA \delta t/2}e^{iB \delta t / 2}e^{iB \delta t/2}e^{iA \delta t} &= e^{i(A + B)\delta t} + O(\delta t^3),
\end{align}
can be used to generate the action of the Hamiltonian $-i\left[A,B\right]$ and the Hamiltonian $A + B$ in the limit $\delta t \rightarrow 0$ \cite{braunsteinQuantumInformationContinuous2005}. By repeated application of the identities above, we can generate evolution which is any superposition of nested commutators of the original set of generators \cite{dalessandroIntroductionQuantumControl2007}.

Starting with the set of generators for $\text{ECD}(\beta)$ and $R_\varphi(\theta)$, $\left\{q \sigma_z, p\sigma_z, \sigma_x, \sigma_y\right\}$,
commutators such as $\left[q \sigma_z, \sigma_x \right] \propto q \sigma_y$ and $\left[\sigma_x, \sigma_y \right] \propto \sigma_z$ can be used to expand the set to $\left\{\sigma_i, q \sigma_i, p \sigma_i\right\}$ where $i \in \left\{x,y,z\right\}$. This shows that effectively, by rotating the qubit between conditional displacements, the ECD gate set can create more general Rabi type interactions between the oscillator and qubit, where qubit-mediated nonlinear gates have been proposed \cite{parkQubitmediatedDeterministicNonlinear2017,parkDeterministicNonlinearPhase2018}.

By using commutators similar to $\left[q\sigma_x, q\sigma_y\right] \propto q^2 \sigma_z$, our set can further be expanded to all quadratic polynomials of $q\sigma_i$ and $p\sigma_i$. This process can be iterated in order to generate any $q^j p^k \sigma_i$ product, where $i \in \left\{x,y,z\right\}$. Terms which do not contain a Pauli operator such as $q^j p^k$ can be generated from commutators such as $\left[q^{j+1}p^{k}\sigma_z, p \sigma_z \right] \propto q^j p^k$. With this, the full Lie algebra for polynomial operators on the qubit and oscillator Hilbert space is generated. 

\subsection*{Speed limit of control}
The maximum interaction rate between the oscillator and qubit will be limited by the maximum displacement in the oscillator before higher-order nonlinear effects begin to invalidate the dispersive approximation. From \cite{blaisCircuitQuantumElectrodynamics2021} the critical oscillator photon number for the $j^\text{th}$ transmon state is
\begin{align}
	n_\text{crit}^j = \frac{1}{2j + 1}\left(\frac{|\Delta - jE_C|^2}{4 g^2} - j\right)
\end{align}
where $\Delta$ is the transmon-oscillator detuning, $g$ is the linear transmon-oscillator coupling rate, and $E_C$ is the charging energy of the transmon. For our experimental parameters, $n_\text{crit}^g \approx 2740$ and $n_\text{crit}^e \approx 910$. These bounds are not strict, however they provide a guiding principle for the maximum photon number before higher order effects become important (see Supplementary Information Section 3.)

With this, the maximum conditional displacement rate is $g_\text{eff}^\text{max} = \alpha_0^\text{max}\chi \approx \sqrt{n_\text{crit}^e}\chi$ using the critical photon number for the first excited state of the transmon. From perturbation theory, the transmon-oscillator dispersive coupling is $\chi \approx \left(2 g^2 E_C\right)/\left(\Delta (\Delta- E_C)\right)$ and the transmon anharmonicity is $K \approx E_C$ \cite{blaisCircuitQuantumElectrodynamics2021}. In the regime $\Delta \gg E_C$ we can approximate $\Delta - E_C \approx \Delta$ and combine the above expressions to find
\begin{align}
	g_\text{eff}^\text{max} \approx \sqrt{\frac{\chi K}{6}}
\end{align}
We note other experiments using sideband three-wave-mixing interactions are similarly limited by a bound $\propto \sqrt{\chi K}$ \cite{pechalMicrowaveControlledGenerationShaped2014,touzardGatedConditionalDisplacement2019,rosenblumCNOTGateMultiphoton2018b}. This suggests that at a fixed dispersive shift, increasing transmon anharmonicity could lead to faster interaction rates, giving a path forward for engineering higher-fidelity gates with enhanced effective three-wave interactions.

\section*{Acknowledgments}
We thank Nicholas Frattini, Rodrigo Cortiñas, Christa Flühmann, and Xu Xiao for helpful discussions on oscillator control and driven nonlinear systems. We are grateful to Jacob Curtis and Billy Kalfus for technical assistance. We thank Ioannis Tsioutsios and Luigi Frunzio for device fabrication assistance. We thank Max Hays, Ben Brock, James Teoh, Chris Wang, Aniket Maiti, Philippe Campagne-Ibarcq, Steven Touzard, and Serge Rosenblum for helpful feedback. We thank the Yale Center for Research Computing for technical support and high performance computing resources. This research was sponsored by the Army Research Office (ARO) under grant number W911NF-18-1-0212, W911NF-16-1-0349, W911NF-18-1-0020, and by the Air Force Office of Scientific Research under grant number FA9550-19-1-0399. The views and conclusions contained in this document are those of the authors and should not be interpreted as representing the official policies, either expressed or implied, of the Army Research Office (ARO), or the U.S. Government. The U.S. Government is authorized to reproduce and distribute reprints for Government purposes notwithstanding any copyright notation herein.

\section*{Author contributions} 
A.E., S.S.E., M.H.D. and R.J.S. developed the large displacement control method. A.E., S.R.J. V.S., and A.Z.D. implemented the numerical ECD parameter optimization. A.E., V.S., and A.Z.D. conducted the measurements. A.E., B.R.,V.S., and S.M.G. developed the theory. J.V. and A.E. performed numerical analysis of the strongly driven nonlinear oscillator. A.E. and M.H.D. wrote the manuscript with feedback from all authors.

\section*{Competing interests}
R.J.S. and M.H.D. are founders and R.J.S. is a shareholder of Quantum Circuits, Inc.

\let\oldaddcontentsline\addcontentsline
\renewcommand{\addcontentsline}[3]{}
\putbib[ECD]
\let\addcontentsline\oldaddcontentsline
\end{bibunit}
\setcounter{equation}{0}
\setcounter{figure}{0}
\setcounter{table}{0}
\setcounter{section}{0}
\makeatletter
\renewcommand{\theequation}{S\arabic{equation}}
\renewcommand{\thefigure}{S\arabic{figure}}
\renewcommand{\thetable}{S\arabic{table}}
\renewcommand{\thesection}{S\arabic{section}}
\crefname{table}{Table}{Tables}
\Crefname{table}{Table}{Tables}
\crefname{figure}{Fig.}{figs.}
\Crefname{figure}{Fig.}{Figures}
\crefname{section}{section}{sections}
\Crefname{section}{Section}{Sections}
\crefname{subsection}{section}{sections}
\Crefname{subsection}{Section}{Sections}

\title{Supplementary Information: Fast Universal Control of an Oscillator with Weak Dispersive Coupling to a Qubit}

\clearpage
\maketitle

{
	\onecolumngrid
	\hypersetup{linkcolor=black}
	\tableofcontents
}

\begin{bibunit}[naturemag]
\section{Experimental design}

\label{sec:exp}
The sample consists of two coaxial microwave cavities machined out of aluminum 6061 alloy anchored at the base stage of a dilution refrigerator operating at $\SI{20}{mK}$. The lower-frequency cavity is used as a high-Q storage oscillator, while the other is overcoupled to a transmission line and used for readout of a fixed-frequency transmon qubit bridging the two cavities. The transmon includes a double-angle-evaporated $\mathrm{Al/AlO_x/Al}$ Josephson-junction fabricated on a sapphire substrate. An FPGA system is used to control the transmon and cavity with a DAC sampling rate of $\SI{1}{GS/s}$. The package and transmon used here is the same as was used in \cite{campagne-ibarcqQuantumErrorCorrection2020}, with device parameters that have aged since that publication. We refer the reader to \cite{campagne-ibarcqQuantumErrorCorrection2020} and associated supplementary material for more details, as well as the wiring diagram, for which the only major difference here is the lowering of amplification power and addition of room-temperature microwave switches on the storage and readout line for better noise properties.

Single-shot readout is performed using a SNAIL parametric amplifier operating with 20 dB of gain in phase-preserving mode \cite{frattiniOptimizingNonlinearityDissipation2018}. We use a square readout pulse of length $\SI{100}{ns}$ and acquire signal for $\SI{824}{ns}$. With additional FPGA delays, the total readout time is $\SI{1.176}{us}$, leading to a readout fidelity greater than $98\%$ as inferred by the measured average contrast of thresholded Rabi fringes. A measurement based feedback routine is used to prepare the transmon in $\ket{g}$ before each experimental iteration.

In equilibrium, the transmon's excited state population is $n_\text{th,q} \sim 0.0092$, corresponding to a temperature of $\sim\SI{68}{mK}$. We rely on a wait time longer than $5T_{1,c}$ between each experiment for the cavity to relax to near-equilibrium. As a conservative estimate, we assume the cavity mode is at the same temperature as the transmon when estimating error sources in \cref{sec:error budget}, corresponding to a cavity excited state population of  $n_\text{th,c} \sim 0.025$ before the start of each experiment. 

\section{System Hamiltonian and Parameters}

Given the range of displacements used in this work, our system is well described by the effective Hamiltonian \cite{niggBlackBoxSuperconductingCircuit2012, minevEnergyparticipationQuantizationJosephson2021a, blaisCircuitQuantumElectrodynamics2021}
\begin{align}
	\label{eq:system Hamiltonian}
	&\frac{H}{\hbar} = \Delta a^\dagger a - \chi a^\dagger a q^\dagger q - \chi'a^{\dagger2} a^2q^\dagger q -K_c a^{\dagger2} a^2 - K_q q^{\dagger2} q^2 + \varepsilon^*(t) a + \Omega^*(t)q + h.c.
\end{align}
where $a$ ($q$) are bosonic annihilation operators for the hybridized oscillator-like (transmon-like) modes. $\Omega(t)$ and $\varepsilon(t)$ are complex-valued drives generated by IQ modulation, and we have ignored terms rotating at twice the drive frequencies. $H$ is written in the co-rotating frame of the qubit and cavity drives, and in this work we use $\Delta = \chi/2$ when performing ECD gates, representing a cavity drive at frequency $(\omega_g + \omega_e)/2$, where $\omega_g$ ($\omega_e$) is the cavity frequency with the transmon in the ground (excited) state. Under this choice, the Hamiltonian given in the main text corresponds to projecting \cref{eq:system Hamiltonian} onto the ground and excited state transmon manifold using the mapping $\sigma_z = 1 - 2q^\dagger q$ and only keeping the dispersive interaction term. Hamiltonian parameters, as well as measured decoherence rates and mode frequencies, are given in \cref{tab:system parameters}, and measurement techniques for these some of these values are described in \cref{sec:calibration}.
\begin{table}[h]
	\begin{tabular}{l@{\hskip 0.25in}l@{\hskip 0.25in}l}
		\textbf{parameter}  & \textbf{value}                                                                \\
		\hline
		transmon g-e transition frequency & $\omega_{ge} = 2\pi \times \SI{6.65}{GHz}$ \\ 
		transmon anharmonicity & $K = 2K_q = 2\pi \times  \SI{193}{MHz}$ \\
		transmon Ramsey coherence & $T_{2R,q} = \SI{30}{us}$ \\
		transmon echo coherence & $T_{2E,q} = \SI{65}{us}$ \\
		bare transmon relaxation&  $ T_{1,q} = \SI{50}{us}$ \\
		transmon relaxation with $\bar{n}_\text{cav} = 900$ &  $ \tilde{T}_{1,q} = \SI{30}{us}$ \\
		transmon equilibrium population & $ n_\text{th} = 0.0092$ \\
		readout frequency & $\omega_r = 2\pi \times \SI{8.22}{GHz}$ \\
		readout dispersive shift & $\chi_{qr} = 2\pi \times\SI{0.96}{MHz}$ \\
		readout relaxation rate & $\kappa_r = \SI{1.7}{MHz}$ \\
		storage cavity frequency & $\omega_c = 2\pi \times \SI{5.26}{GHz}$\\
		storage dispersive shift  & $\chi = 2\pi \times \SI{32.8}{kHz}$ \\
		storage second-order dispersive shift & $\chi' = 2\pi \times \SI{1.5}{Hz}$\\
		storage cavity Kerr & $2K_c = 2\pi \times \SI{1}{Hz}$\\
		storage cavity relaxation & $T_{1,c} = \SI{436}{us}$ \\
		storage cavity Ramsey coherence & $T_{2R,c} < 2 T_{1,c} = \SI{872}{us}$
	\end{tabular}
	\caption{\label{tab:system parameters} \textbf{Measured system parameters and loss rates}. Measurement of the dispersive shift, the second order dispersive shift, and Kerr is described in \cref{sec:hamiltonian measurement}.  The Cavity relaxation time $T_{1,c}$ is measured by preparing a coherent state $\alpha_0 = 3.6$ and measuring $\braket{\hat{a}^\dagger \hat{a}}(t)$ using time-dependent transmon spectroscopy. The limit on the cavity Ramsey coherence time $T_{2,c}$ is inferred from the cavity relaxation time.}
\end{table}

\section{System Characterization and Calibration}

\label{sec:calibration}
In this section, we outline techniques to characterize an oscillator and qubit coupled with $\chi/2\pi$ on the order of or smaller than qubit decoherence rates. In this regime, methods for calibration of control and Hamiltonian parameters which rely on large number-splitting are inefficient. For these calibrations, we rely on semiclassical phase-space trajectories as derived in \cref{sec:trajectories}. In \cref{sec:out and back}, we describe the \textit{out-and-back} method, which uses large displacements of the oscillator mode to realize a measurement of Hamiltonian parameters. Finally in \cref{sec:geometric phase} we describe a simple geometric phase measurement which is used to calibrate the oscillator drive strength $|\varepsilon|$.

\subsection{Semiclassical phase space trajectories}

\label{sec:trajectories}
Starting from $H$ in \cref{eq:system Hamiltonian}, we perform a time-dependent displaced frame transformation using the unitary $U = D^\dag(\alpha(t)) = \exp\left\{\alpha^*(t) a - \alpha(t) a^\dagger\right\}$. This modifies the state according to $\tilde{\rho}(t) = D^\dag(\alpha(t)) \rho(t) D(\alpha(t))$, and the Hamiltonian according to
$H \rightarrow \tilde{H} = UH U^\dagger + \left(i\hbar\right) \left(\partial_t U\right) U^\dagger = D^\dagger \left(\alpha(t)\right) H D\left(\alpha(t)\right) + \left(i\hbar\right) \left(a \partial_t \alpha^*(t) - a^\dagger \partial_t \alpha(t)\right)$, giving
\begin{equation}
	\label{eq:displaced H}
	\begin{aligned}
		\frac{\tilde{H}}{\hbar} &=  \Delta a^\dagger a - (\chi + 4\chi'|\alpha|^2) a^\dagger a q^\dagger q - \chi' a^{\dagger2} a^2 q^\dagger q - K_c a^{\dagger2} a^2 - K_q q^{\dagger 2}q^2 - (\chi + 2 |\alpha|^2 \chi')(\alpha^*a + \alpha a^\dagger)q^\dagger q &  \\
		&- (\chi|\alpha|^2 + \chi' |\alpha|^4)q^\dagger q - 4K_c|\alpha|^2 a^\dagger a + \left(\Delta \alpha^* -2K_c|\alpha|^2\alpha^* + i (\partial_t \alpha^*) + i \frac{\kappa}{2} \alpha^* + \varepsilon^* \right)a + h.c. &  \\
		&- K_c\left(2\alpha a^{\dagger 2}a + \alpha^2 a^{\dagger2} + h.c.\right) - \chi'\left(2\alpha a^{\dagger 2}a + \alpha^2 a^{\dagger2} + h.c.\right)q^\dagger q+ \Omega^*(t)q + h.c.
	\end{aligned}
\end{equation}
We have also included the deterministic part of oscillator relaxation at a rate $\kappa/2$ as described in the methods section of the main text.

\subsubsection{Simulating in the displaced frame}

\label{sec:simulation displaced frame}
Given an oscillator drive $\varepsilon(t)$, it is numerically advantageous to cancel the classical part of the oscillator's phase-space trajectory by picking a time-dependent frame $\alpha(t)$ which cancels the displacement term in $\tilde{H}$. This is done by solving
\begin{equation}
	\label{eq:cm trajectory}
	\begin{aligned}
		\partial_t \alpha(t) &= -i \Delta \alpha(t) + 2i K_c |\alpha(t)|^2 \alpha(t) - \frac{\kappa}{2} \alpha(t) - i \varepsilon(t) \\
		\alpha(0) &= 0
	\end{aligned}
\end{equation}
for $\alpha(t)$. Once $\alpha(t)$ is known, $\tilde{H}$ without the linear terms in $a$ and $a^\dagger$ is used to efficiently simulate a cavity and transmon evolution in the displaced frame using a truncated Hilbert space. This displaced frame Hamiltonian is used for master equation simulations in \cref{sec:error budget}.

\subsubsection{Semiclassical trajectories}

During a period where the qubit populations stay constant ($\Omega(t) =0$), we can instead determine the oscillator's phase-space trajectories conditioned on the qubit's ground or excited state. This semiclassical approximation is done by replacing $q^\dagger q$ in $\tilde{H}$ with $\left\{0,1\right\}$ for transmon states $\left\{\ket{g},\ket{e}\right\}$. This replacement gives the two Hamiltonian sectors
\begin{equation}
	\begin{aligned}
		\frac{\tilde{H}_g}{\hbar} &=  (\Delta - 4K_c|\alpha|^2) a^\dagger a - K_c a^{\dagger2} a^2 + \left(\Delta \alpha^* -2K_c|\alpha|^2\alpha^* + i (\partial_t \alpha^*) + i \frac{\kappa}{2} \alpha^* + \varepsilon^* \right)a + h.c.  \\
		&- K_c\left(2\alpha a^{\dagger 2}a + \alpha^2 a^{\dagger2} + h.c.\right), \\
		\frac{\tilde{H}_e}{\hbar} &=  (\Delta - \chi - 4\chi'|\alpha|^2 - 4K_c|\alpha|^2) a^\dagger a - (\chi' + K_c) a^{\dagger2} a^2 \\
		&+ \left(\Delta \alpha^* -2K_c|\alpha|^2\alpha^* + i (\partial_t \alpha^*) + i \frac{\kappa}{2} \alpha^* + \varepsilon^* - (\chi + 2 |\alpha|^2 \chi')\alpha^* \right)a + h.c.  \\
		&- (K_c + \chi')\left(2\alpha a^{\dagger 2}a + \alpha^2 a^{\dagger2} + h.c.\right)
	\end{aligned}
\end{equation}
describing the dynamics of the driven oscillator when the transmon is in the ground or excited state. Similar to the displaced-frame simulation, the linear part of these Hamiltonians can be individually cancelled, resulting in the two equations
\begin{equation}
	\label{eq:nonlinear trajectories}
	\begin{aligned}
		\partial_t \alpha_g(t) &= -i \Delta \alpha_g(t) + 2i K_c |\alpha_g(t)|^2 \alpha_g(t) - \frac{\kappa}{2} \alpha_g(t) - i \varepsilon(t)\\
		\partial_t \alpha_e(t) &=-i \Delta \alpha_e(t) + 2i K_c |\alpha_e(t)|^2 \alpha_e(t) - \frac{\kappa}{2} \alpha_e(t) - i \varepsilon(t) + i(\chi + 2\chi'|\alpha_e(t)|^2)\alpha_e(t)
	\end{aligned}
\end{equation}
which can be used to calculate the semiclassical trajectories for the ground or excited states during periods when $\Omega(t) = 0$. In the case of a conditional displacement, after each $\pi$ pulse, the Hamiltonians are swapped, and the result from the previous part of the trajectory is used to seed the next initial value problem. In our simulations, we solve these nonlinear initial value problems using a central-difference method with trajectories sampled at $\SI{1}{ns}$. These trajectories are used in \cref{sec:hamiltonian measurement} for Hamiltonian parameter calibration and in \cref{sec:ECD optimization} for optimization of the cavity and qubit drives to produce ECD gates.

\subsection{Measurement of Hamiltonian parameters using out-and-back}

\label{sec:hamiltonian measurement}
\label{sec:out and back}

\begin{figure*}[!ht]
	\begin{center}
		\vspace{-2\baselineskip}
		\phantomsubfloat{\label{fig:out and back a}}
		\phantomsubfloat{\label{fig:out and back b}}
		\phantomsubfloat{\label{fig:out and back c}}
		\phantomsubfloat{\label{fig:out and back d}}
		\includegraphics{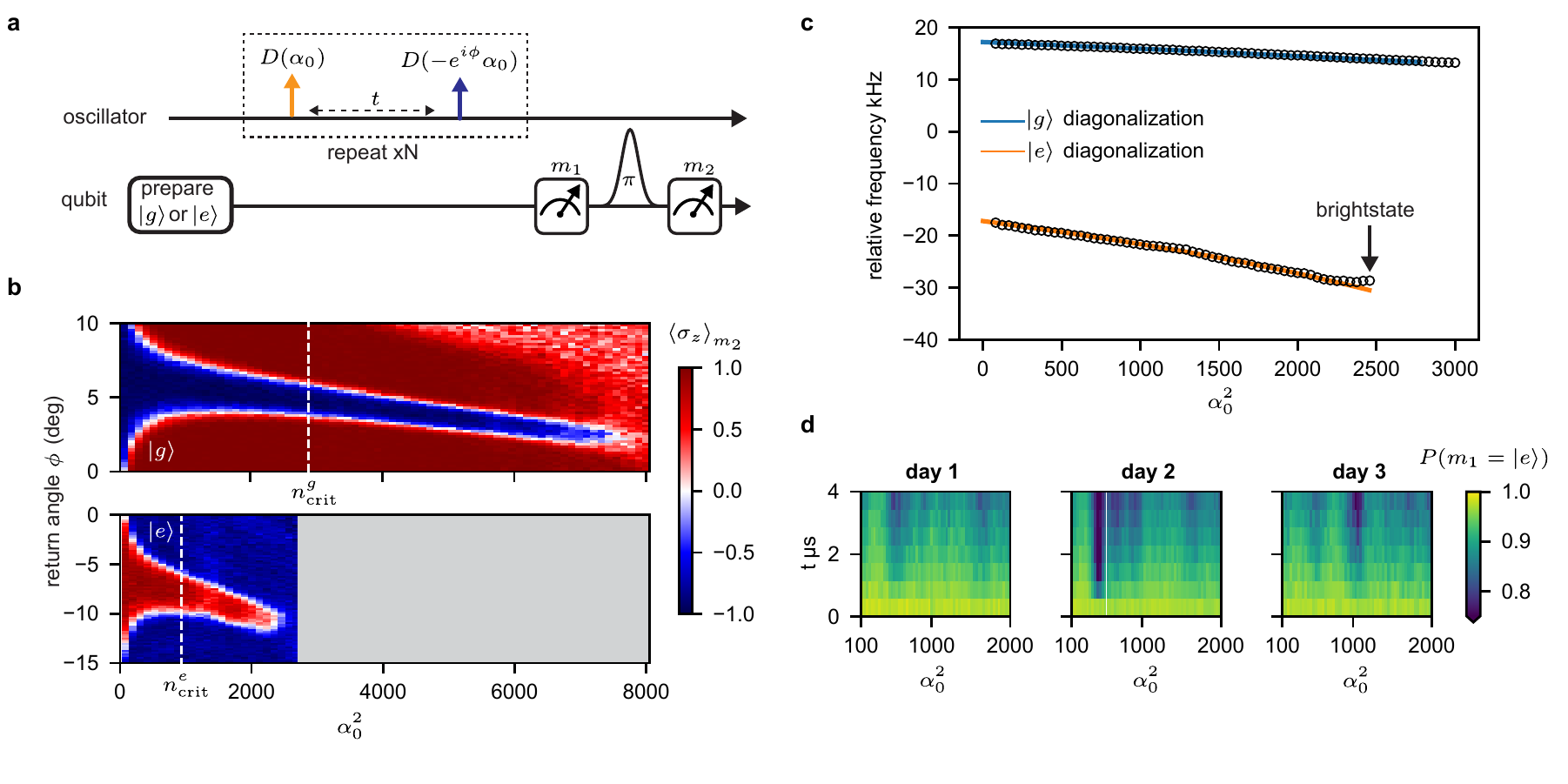}
		\caption{\label{fig:out and back} \textbf{Out-and-back measurement} \textbf{a.} Experimental sequence to measure the phase accumulation of a coherent state with radius $\alpha_0$ after a time $t$ when the transmon is in the ground or excited state. \textbf{b.} Measurement result with initial transmon state $\ket{g}$ (top panel) and $\ket{e}$ (bottom panel) with a fixed wait time of $t=\SI{1}{us}$ and $N=5$ repetitions post-selected on measurement $m_1$ matching the prepared state. \textbf{c.} Measured cavity frequency dispersion (open circles) compared to numerical diagonalization of \cref{eq:coupled H} (solid lines). The open circles is a fit of different experimental data than shown in panel (b), with higher resolution out to $\alpha_0^2 = 3000$. \textbf{d} Measured transmon relaxation as a function of $\braket{n}$ found by analyzing $m_1$ measurement results with initial transmon state $\ket{e}$ and sweeping $t$. $\phi$ is chosen close to values which successfully displace the cavity state back to the origin of phase space. Measurement results shown for three consecutive days with each experiment averaging for around 1 hour.}
	\end{center}
\end{figure*}

Although many of the Hamiltonian terms are small relative to the rate of transmon decoherence, they can be estimated in experiment using large displacements to enhance their effective strength. Here, we make use of this enhancement through the measurement sequence shown in \cref{fig:out and back a}. We first prepare the qubit in the ground or excited state then displace the oscillator out by $\alpha_0$. After a time $t$, the oscillator is displaced back by $-e^{i\phi} \alpha_0$, where $\phi$ is swept. The second displacement serves as an attempt to displace the oscillator's state back to the origin of phase-space. If the attempt is successful, $\phi$ encodes the oscillator's coherent-state phase accumulation at a displacement $\alpha_0$ after a time $t$. A narrow-bandwidth $\pi$ pulse ($\sigma = \SI{200}{ns}$) is then used as a probe, only flipping the transmon's state measured by $m_2$ if the oscillator's state is close to the origin of phase space. Finally, we postselect the results of $m_2$ on the condition $m_1 = \ket{\psi_i}$, where $\ket{\psi_i}$ is the initial transmon state, in order to remove the influence of transmon relaxation or heating. 

In \cref{fig:out and back c}, we show the results of this experiment with $t = \SI{1}{us}$ and $N = 5$ repetitions of out and back, used to enhance the sensitivity. In initial state $\ket{e}$, the signal is lost above $\braket{n}\gtrsim 2500$ photons, and in initial state $\ket{g}$, the signal is lost above $\braket{n} \gtrsim 7500$ photons. These values represent the oscillator photon numbers at which the qubit is excited outside of the $\ket{g},\ket{e}$ manifold due to higher-order nonlinear transitions, a process sometimes referred to as \textit{bright-stating} that has been observed in previous experiments using readout resonators \cite{reedHighFidelityReadoutCircuit2010a, sankMeasurementInducedStateTransitions2016, lescanneEscapeDrivenQuantum2019}. Such an effect could be suppressed by using an inductive shunt, proving a path forward to engineering faster gates \cite{verneyStructuralInstabilityDriven2019a}. \cref{fig:out and back} also indicates the critical photon number at which the dispersive approximation begins to fail calculated using the expressions in the methods section of the main text, $n_\text{crit}^g \approx 2740$ and $n_\text{crit}^e \approx 910$. This serves as a guiding principle for the speed limit of control.

At each $\braket{n}$ below the bright-stating point, the relative oscillator frequency when the transmon is in the ground or excited state is obtained by fitting $\braket{\sigma_z}$ to a Gaussian function and dividing its mean phase accumulation by the wait time $t$. The resulting relative frequency dispersion as a function of average cavity photon number $\braket{n} = \alpha_0^2$ for transmon states $\ket{g}$ and $\ket{e}$ is shown in \cref{fig:out and back c} by the open circles. Note that the finite duration of the displacement pulses will also influence the result, and as a secondary check, the experiment can be repeated at different wait times $t$. Clearly the dispersion is not constant, representing the need to include higher-order nonlinear terms beyond the dispersive coupling in our effective Hamiltonian, as is done in \cref{eq:system Hamiltonian}.

To extract the effective Hamiltonian parameters, we fit the measured dispersion (open circles in \cref{fig:out and back}) to that expected from $H$ in \cref{eq:system Hamiltonian}. In particular, the semiclassical trajectories in \cref{eq:nonlinear trajectories} which govern the evolution of coherent states predict the effective cavity rotation frequencies to be 
\begin{align}
	\Delta_{g} &= \Delta - 2K_c\alpha_0^2\\
	\Delta_{e} &= \Delta - \chi -  (2K_c + 2\chi')\alpha_0^2 
\end{align}
as a function of the average number of photons in the cavity $\alpha_0^2$ when the transmon is in the ground and excited states respectively. By fitting the sum and difference of two dispersion curves to the sum and difference of these linear functions, we can extract the four unknown Hamiltonian parameters $\chi$, $\chi'$, $K$, and $\Delta$. The dispersion fits well to a linear function in the range of interest for control, for photon numbers up to $\alpha_0^2 < 2000$, with results given in \cref{tab:system parameters}. For experiments using echoed conditional displacements, we use this experiment to calibrate the cavity drive frequency such that $\Delta = \chi/2$.

To further confirm our model, we compare the measured dispersion to a numerical diagonalization of the coupled transmon-oscillator Hamiltonian 
\begin{equation}
	\label{eq:coupled H}
	\frac{H}{\hbar} = 4 E_c (\hat{N} - N_g)^2 - E_J \cos\left(\hat{\varphi}\right) +  \omega_{c}^{\text{bare}} a^\dagger a + g (\hat{N} - N_g)\left(a + a^\dagger \right)
\end{equation}
where $\hat{N}$ is the Cooper-pair number operator, $\hat{\varphi}$ is the conjugate Josephson phase, $N_g$ is the offset charge in units of $2e$, and $\hat{a}$ is the bare cavity mode annihilation operator \cite{kochChargeinsensitiveQubitDesign2007}. For photon numbers up to $\braket{n} \approx 1000$, we have confirmed that the oscillator's dispersion for qubit states $\ket{g}$ and $\ket{e}$ does not depend on offset charge, so we set $N_g = 0$. With this choice, we use second-order perturbation theory to find the bare Hamiltonian parameters which fit the measured hybridized mode frequencies, transmon anharmonicity, and dispersion at low $\braket{n} \approx 0$, resulting in the bare parameters $g/2\pi = \SI{9.12}{MHz}$, $E_j/2\pi = \SI{32.33}{GHz}$, $E_c/2\pi = \SI{181}{MHz}$ and $\omega_{c}^{\text{bare}} = \SI{5.26}{GHz}$. Using these parameters, we numerically diagonalize \cref{eq:coupled H} in the basis of transmon eigenstates with a Hilbert space of 2800 oscillator states and 12 transmon states. The resulting dispersion is shown by the solid lines in \cref{fig:out and back c}, which have excellent agreement with the measured dispersion. Also, the diagonalization predicts a breakdown of our quantum number assignment algorithm for transmon state $\ket{e}$ at 2500 photons in the oscillator, matching the measured breakdown in experiment. Finally, we note that higher-order perturbation theory can also be used to predict an analytic effective Hamiltonian at large photon numbers in the cavity, as shown in \cite{venkatramanStaticEffectiveHamiltonian2022}.

We use a similar out-and-back measurement to probe the transmon relaxation and heating rate while the cavity is displaced to a large coherent state, since a reduction in transmon lifetime has been observed when displacing readout resonators \cite{minevCatchReverseQuantum2019}. For this, we use the out-and-back sequence in \cref{fig:out and back a}, except we sweep $t$, and fix $\phi(t)$ close to phases that displace the oscillator's state back to the origin of phase-space at each $\braket{n}$ given the measured dispersion. In this case, we focus on the result of $m_1$. We find that, up to $2000$ oscillator photons, there is no appreciable heating out of $\ket{g}$ when displacing the cavity state, indicating that the dressed dephasing rate is small \cite{boissonneaultDispersiveRegimeCircuit2009}. However, when preparing the transmon in $\ket{e}$, we measure that the transmon's relaxation rate shows a dependence on cavity photon number. In \cref{fig:out and back d}, we plot the measured probability of the transmon remaining in $\ket{e}$ after a wait time $t$ up to $\SI{4}{us}$ when displacing the cavity to $\braket{n} = \alpha_0^2$, with experiments run on three consecutive days. We suspect this time-dependent $\tilde{T}_{1,q}$ vs $\bar{n}_\text{cav}$ effect is caused by fluctuating two-level-systems (TLS) which come into resonance with the transmon as it is stark-shifted by cavity photons \cite{klimovFluctuationsEnergyRelaxationTimes2018, burnettDecoherenceBenchmarkingSuperconducting2019,carrollDynamicsSuperconductingQubit2021}. Although the $T_1$ vs $\bar{n}$ changes with time, we find an average value of $\tilde{T}_{1,q} \approx \SI{30}{us}$ for $\bar{n} = 900$ over the data plotted in \cref{fig:out and back d}, and we use this value when performing master equation simulations in \cref{sec:error budget}.

\subsection{Oscillator drive strength calibration using geometric phase}

\label{sec:geometric phase}
\begin{figure*}[ht]
	\begin{center}
		\vspace{-2\baselineskip}
		\phantomsubfloat{\label{fig:geometric phase a}}
		\phantomsubfloat{\label{fig:geometric phase b}}
		\phantomsubfloat{\label{fig:geometric phase c}}
		\includegraphics{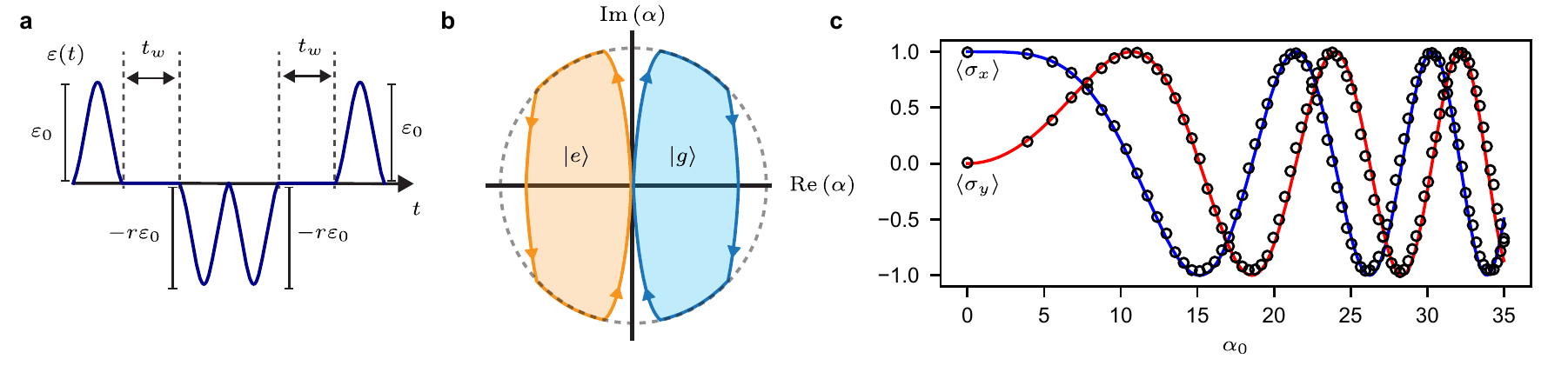}
		\caption{\label{fig:geometric phase} \textbf{Geometric phase measurement} \textbf{a.} Geometric phase measurement drive sequence and drive parameterization. \textbf{b.} Phase space trajectories for $\alpha_g$ and $\alpha_e$. \textbf{c} Measured $\braket{\sigma_x}$ and $\braket{\sigma_y}$ compared to expected phase found by solving the semiclassical trajectories in \cref{eq:nonlinear trajectories} (solid lines) as a function of phase space radius $\alpha_0$.}
	\end{center}
\end{figure*}

In this section, we discuss a simple experiment which can be used to calibrate a linear oscillator drive amplitude in the weak-$\chi$ regime. Starting with the qubit prepared in $\ket{\psi_i} = \frac{1}{\sqrt{2}}\left(\ket{g} + \ket{e}\right)$, we construct an oscillator drive sequence which encloses an area in phase space for both $\ket{g}$ and $\ket{e}$ trajectories. By disentangling the qubit and oscillator at the end of the sequence, the qubit will be left in the state $\ket{\psi_f} = \frac{1}{\sqrt{2}}\left(\ket{g} + e^{i\phi}\ket{e}\right)$ where $\phi$ encodes the enclosed area \cite{chaturvediBerryPhaseCoherent1987,vacantiGeometricphaseBackactionMesoscopic2012,pechalGeometricPhaseNonadiabatic2012a,songContinuousvariableGeometricPhase2017}. Given an arbitrary displacement pulse shape $g(t)$ with a length $t_p$, a simple pulse sequence that accomplishes this is 
\begin{equation}
	\varepsilon(t) = \varepsilon_0\left[g(t) - r g(t - (t_p + t_w)) - r g(t - (2t_p + t_w)) + g(t - (2t_p + 2t_w))\right]
\end{equation}
as shown in \cref{fig:geometric phase a} with a phase-space trajectory shown in \cref{fig:geometric phase b}. This drive is similar to a conditional displacement without a qubit $\pi$ pulse, and the goal here is to calibrate the pulse scale $\varepsilon_0$.

To analyze this sequence, we note that for low photon numbers, the Hamiltonian is well described by only the dispersive term. With this approximation, the semiclassical trajectories in \cref{eq:nonlinear trajectories} can be solved with initial value $\alpha_{_{e}^{g}}(0) = 0$ giving
\begin{equation}
	\label{eq:dispersive ivp sol}
	\alpha_{_{e}^{g}}(t) = e^{-\frac{1}{2}\left(\pm i\chi + \kappa\right)t} \left(e^{\frac{1}{2}\left(\pm i\chi + \kappa\right)t_0}\alpha_{_{e}^{g}}(t
	_0) - i \int_{t_0}^t e^{\frac{1}{2}\left(\pm i\chi + \kappa\right)\tau}\varepsilon(\tau)d\tau\right).
\end{equation}

Subsituting $\varepsilon(t)$ into \cref{eq:dispersive ivp sol}, we solve for the ratio of the middle-two pulses, $r$, such that the condition $\alpha_{_e^g} = 0$ is satisfied at the end of the entire sequence, and the qubit and oscillator are disentangled. Remarkably, this ratio is independent of the shape of the displacement pulse $g(t)$, and is found to be
\begin{equation}
	\label{eq:geometric pulse ratio}
	r = \frac{1 + e^{\frac{1}{2}\left(\pm i\chi + \kappa\right)(3t_p + 2t_w)} }{e^{\frac{1}{2}\left(\pm i\chi + \kappa\right)(t_p + t_w)} + e^{\frac{1}{2}\left(\pm i\chi + \kappa\right)(2t_p + t_w)} }
\end{equation}
where $t_p$ is the length of the displacement pulses $g(t)$ and $t_w$ is the wait time. By Taylor expanding this ratio in orders of $\kappa$, we find in the limit $(t_p + t_w) << 1/\kappa$,
\begin{equation}
	r = \cos\left(\frac{\chi}{4}(3 t_p + 2t_w)\right)\sec\left(\frac{\chi}{4} t_p\right)
\end{equation}
independent of the qubit state. In the high-Q oscillator limit, this sequence disentangles the oscillator and qubit and can be used to measure the geometric area enclosed by the trajectory.

Using \cref{eq:dispersive ivp sol}, we numerically integrate the sequence with displacement pulses $g(t)$ chosen as truncated Gaussians with $\sigma = \SI{11}{ns}$ and $t_p = \SI{44}{ns}$ to find the phase-space trajectories for $\alpha_{_{e}^{g}}$, including the maximum phase-space radius $\alpha_0$, and the associated geometric phase difference. By measuring $\braket{\sigma_x}$ and $\braket{\sigma_y}$ while sweeping the drive scale, this phase difference is fit to the experiment, allowing a calibration of $\varepsilon_0$ in terms of DAC amplitude. In \cref{fig:geometric phase c} we show an example of the measured phase accumulation as $\varepsilon_0$ is linearly scaled ($t_w = \SI{200}{ns}$), along with the phase predicted using the integrated geometric area. If desired, nonlinear effects can be included by using \cref{eq:nonlinear trajectories} when calculating the trajectories, however in this case it is no longer guaranteed that $r$ given by \cref{eq:geometric pulse ratio} will exactly disentangle the oscillator and qubit. Using this, we find a maximum drive amplitude of $|\varepsilon|_\text{max}/2\pi \approx \SI{400}{MHz}$ before saturating our room-temperature amplification chain. 

\section{The Echoed Conditional Displacement Gate}

In \cref{sec:ideal gate} we derive the echoed conditional displacement unitary assuming an ideal dispersive Hamiltonian. Next, in \cref{sec:ECD optimization}, we optimize the ECD gate considering drive constraints and  higher-order nonlinearities.

\subsection{Derivation of the ECD gate}

\label{sec:ideal gate}
Here, $\varepsilon(t)$ is a complex-valued function representing the envelope of an I-Q modulated drive with a carrier frequency $(\omega_g + \omega_e)/2$, where $\omega_g$ ($\omega_e$) is the oscillator's frequency when the qubit is in the ground (excited) state. In the co-rotating frame at the drive frequency and qubit frequency, the ideal Hamiltonian is
\begin{equation}
	\frac{H}{\hbar} = \chi a^\dagger a \frac{\sigma_z}{2} + \varepsilon^*(t) a + \varepsilon(t) a^\dagger
\end{equation}
where we have neglected terms rotating at twice the drive frequency.

The echoed conditional displacement gate consists of two driving steps with a qubit $\pi$ pulse between, here assumed instantaneous. In this case, the general solution to the time-dependent Schrödinger equation $i\hbar\partial_t U = H U$ is
\begin{equation}
	U = \mathcal{T} e^{-\frac{i}{\hbar} \int_{t_1}^{T} H(\tau) d\tau}\sigma_x \mathcal{T}e^{-\frac{i}{\hbar} \int_{0}^{t_1} H(\tau) d\tau}
\end{equation}
where $t_1$ is the time of the $\pi$ pulse, $T$ is the total time of the gate, and $\mathcal{T}$ is the time-ordering operator.

To represent the action of the $\pi$ pulse flipping the sign of $\sigma_z$ between the two trajectories, we instead modify the dispersive Hamiltonian to include a function $z(t) = \pm 1$ which represents the sign of $\sigma_z$, giving
\begin{equation}
	\frac{H}{\hbar} = \chi a^\dagger a \frac{\sigma_z z(t)}{2} + \varepsilon^*(t) a + \varepsilon(t) a^\dagger.
\end{equation}

Using this modified Hamiltonian, we take as an ansatz for the solution of the Schrödinger equation
\begin{equation}
	U = e^{i \theta \frac{\sigma_z}{2}}e^{a^\dagger (\gamma + \delta \sigma_z) - a(\gamma^* + \delta^* \sigma_z)}e^{i\phi a^\dagger a \sigma_z}
\end{equation}
where $\theta$ represents an ancilla qubit phase, $\gamma$ and $\delta$ represent a displacement and conditional displacement, $\phi$ represents a qubit state-dependent rotation of the oscillator, and these variables are time-dependent. 

Ignoring a global phase, the Schrödinger equation gives
\begin{align}
	\partial_t \theta &= - 2 \text{Re}\left[\delta \varepsilon^*\right] \\
	\partial_t \gamma &= -i\frac{\chi}{2} z(t)\delta - i\varepsilon \\
	\partial_t \delta &= -i \frac{\chi}{2}z(t) \gamma \\
	\partial_t \phi &= -\frac{\chi}{2}z(t).
\end{align}
These equations can be solved, giving
\begin{align}
	\label{eq:sol start}
	\theta(t) &= -2 \int_{0}^{t} d\tau \text{Re}\left[\varepsilon^*(\tau)\delta(\tau)\right] \\
	\gamma(t) &= -i \int_{0}^{t} d\tau \cos\left[\phi(\tau) - \phi(t)\right]\varepsilon(\tau) \\
	\delta(t) &= -\int_{0}^{t} d\tau \sin\left[\phi(\tau) - \phi(t)\right]\varepsilon(\tau) \\
	\phi(t) &= -\frac{\chi}{2} \int_{0}^{t} d\tau z(\tau).
	\label{eq:sol end}
\end{align}
The state-dependent rotation of the oscillator can be canceled by setting $\phi(T) = 0$. For the echoed conditional displacements in this work, this is done by applying a single qubit $\pi$ pulse at time $T/2$. If desired, more qubit echos can be included, subject to the condition $\int_{0}^{T} d\tau z(\tau) =0$.

Using the Baker-Campbell-Hausdorff formula, the conditional displacement and displacement can be separated, giving the overall unitary
\begin{equation}
	\label{eq:cd_unitary}
	U = \sigma_xe^{i \theta' \frac{\sigma_z}{2}} D(\lambda) CD(\beta)
\end{equation}
corresponding to a conditional displacement $CD(\beta) = D(\beta/2)\ket{g}\bra{g} + D(-\beta/2)\ket{e}\bra{e}$, a displacement, and an additional qubit phase with the parameters $\beta = 2\delta(T)$, $\lambda = \gamma(T)$, and $\theta' = \theta(T) + 2 \text{Im}\left[\gamma(T)\delta(T)\right]$. We have also explicitly included a $\sigma_x$ operator to represent the action of the single $\pi$ pulse (here $\text{ECD}\left(\beta\right) = \sigma_x CD(\beta)$).

To realize an ECD gate, we aim to null the qubit phase and oscillator displacement. In the limit of instantaneous displacements and motivated by the geometry of rotating phase-space, this can be perfectly achieved by choosing the drive
\begin{align}
	\label{eq:perfect drive}
	\varepsilon(t) = \alpha\left[\delta(t) - 2\delta(t - T/2)\cos\left(\chi T /4\right) + \delta(t - T)\cos\left(\chi T/2 \right)\right]
\end{align}
where $\delta(t)$ is the Dirac delta function. This drive corresponds to the displacement sequence described in Fig.\ 1b of the main text with $\beta = 2 i \alpha e^{i\phi} \sin\left(\chi T/2\right)$ and $|\alpha| = \alpha_0$. 

The drive in \cref{eq:perfect drive} cannot be realized in experiment due to bandwidth and amplitude limits of a realistic microwave drive.
Also, effects such as photon loss, higher-order nonlinearities, and the finite duration of the qubit $\pi$ pulse are not taken into account in \Crefrange{eq:sol start}{eq:sol end}. 
To realize a high-fidelity ECD gate in the presence of these effects, we optimize $\varepsilon(t)$ using semiclassical trajectories as described in the next \cref{sec:ECD optimization}.

\subsection{Optimization of the ECD gate}

\label{sec:ECD optimization}
In our experiment, the dynamics can slightly differ from those described by \Crefrange{eq:sol start}{eq:sol end} due to the second-order dispersive shift $\chi'$ and the oscillator Kerr $K_c$ which become relevant at large phase-space displacements. These effects can be studied by examining the displaced-frame Hamiltonian in \cref{eq:displaced H}. In $\tilde{H}$, nonlinear terms proportional to $\chi'$ or $K_c$ can generally cause distortions to the state. However, simulations indicate that these terms do not significantly decrease the fidelity of ECD control protocols given our system parameters in \cref{tab:system parameters}. This is partially because the deleterious effect of terms which are proportional to $\text{sign}(\alpha)$ or $\text{sign}(\sigma_z)$ are significantly reduced due to the phase-space echo $\alpha \rightarrow -\alpha$ and qubit echo $\ket{g} \leftrightarrow \ket{e}$ at time $T/2$ during the ECD gate, which cancels part of their on-average effect to the state distortion in the same way the qubit-state dependent oscillator rotation is canceled during the ideal ECD gate. 

With this in mind, we optimize ECD gates using semiclassical trajectories (\cref{sec:trajectories}) which account for the linear displaced-frame terms (proportional to $a$), including those caused by the second-order disperse shift, Kerr, and photon loss. We assume a form of the unitary still given by \cref{eq:cd_unitary}, with the values of $\beta$, $\lambda$ and $\theta'$ calculated using the trajectories $\alpha_{_e^g}(t)$ for the ground and excited qubit states. We note that $\text{ECD}(\beta)$ can be generated with any $\alpha_0$ or $\chi$ as long as $\chi T < 2\pi$ and the qubit rotation pulse bandwidth is sufficiently large compared to $\chi$. In particular, a large $\alpha_0$ is not required.

\begin{figure*}[!ht]
	\begin{center}
		\vspace{-2\baselineskip}
		\phantomsubfloat{\label{fig:CD drives a}}
		\phantomsubfloat{\label{fig:CD drives b}}
		\phantomsubfloat{\label{fig:CD drives c}}
		\includegraphics{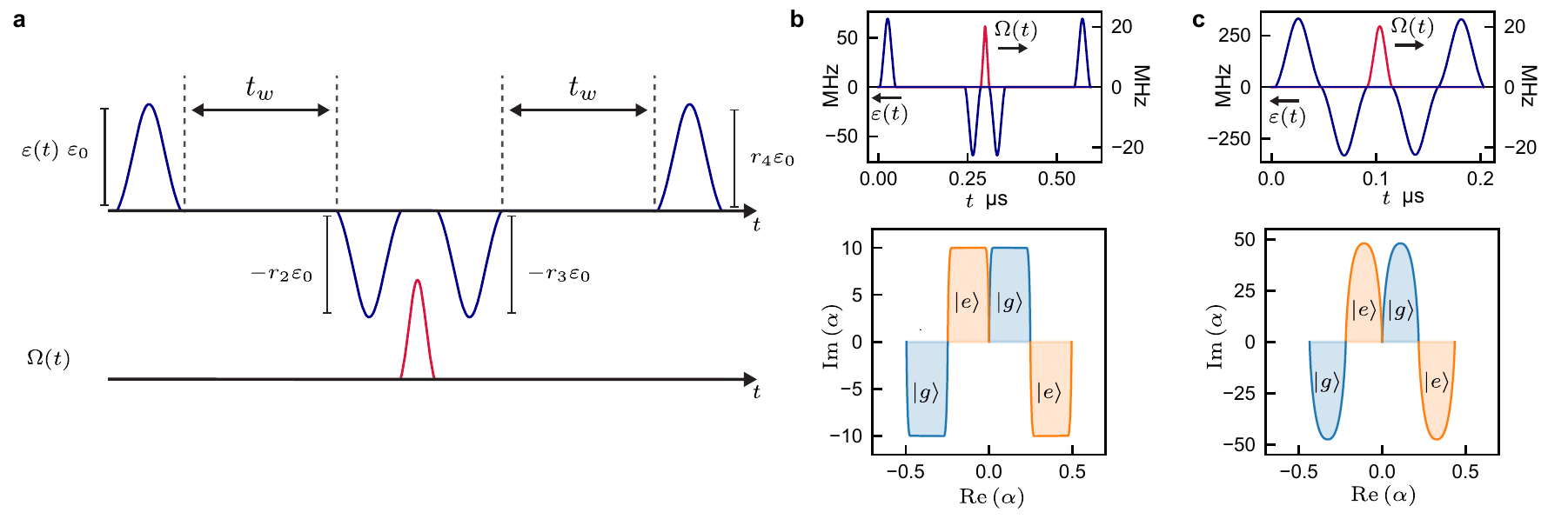}
		\caption{\label{fig:CD drives} \textbf{Echoed conditional displacement} \textbf{a.} Parameterization of the drive $\varepsilon(t)$ for the echoed conditional displacement. \textbf{b. and c.} Echoeded conditional displacement gate for $\beta = 1$ optimized with target $\alpha_0=10$ (\textbf{b}) and $\alpha_0 = 50$ (c). Top panels show the resulting drives, and bottom panels show the semi-classical phase space trajectories at a large aspect ratio.}
	\end{center}
\end{figure*}

To construct each ECD gate, we start by imposing the drive to be of the form shown in \cref{fig:CD drives a}, which replaces the Dirac $\delta$-functions in \cref{eq:perfect drive} with fixed-length Gaussian waveforms, chosen in our experiment with a standard deviation of $\sigma = \SI{11}{ns}$ and a total length of $4\sigma = \SI{44}{ns}$. This simplification is chosen so that the drive strength required to realize the large displacements used in this work remain in the linear regime of our room temperature amplification chain, and so the displacements take the exact same form as those used to calibrate the drive amplitude using the geometric phase measurement shown in \cref{sec:geometric phase}. The amplitude ratio of the second, third, and fourth Gaussian to the first are given by $r_2$, $r_3$, and $r_4$, and the wait time between the displacements is given by $t_w$.

Using a simple optimization strategy, we find the values of $\left\{\varepsilon_0, r_2,r_3, r_4 \right\}$ which realize a target conditional displacement $\beta$ with intermediate phase-space radius $\alpha_0$ in the shortest time $t_w$. Starting with a large guess time $t_w$, the parameters are optimized with a Nelder-Mead method using the cost function
\begin{align}
	\text{C}=
	\left|\alpha_g(T/2) + \alpha_e(T/2)\right|^2 + \left|\alpha_g(T) + \alpha_e(T)\right|^2 + \left(\frac{\left|\alpha_g(T/4)+ \alpha_e(T/4)\right|}{2} - \alpha_0\right)^2 + \left(\frac{\left|\alpha_g(3T/4)+ \alpha_e(3T/4)\right|}{2} - \alpha_0\right)^2 
\end{align}
where $\alpha_{_e^g}(t)$ is calculated using \cref{eq:nonlinear trajectories} including the second-order dispersive shift, Kerr, and photon loss.
This cost function minimizes the final and midpoint net displacement, and ensures an intermediate phase space radius of $\alpha_0$ for the first and third displacements. Once $\left\{\varepsilon_0, r_2,r_3, r_4 \right\}$ have converged at a given $t_w$, $\beta$ is calculated as $\alpha_e(T) - \alpha_g(T)$, and $t_w$ is stepped down until the target $\beta$ is realized. If $t_w$ reaches 0, and the target $\beta$ has not been reached, then $\alpha_0$ is reduced until the target $\beta$ is realized. We note that shorter pulses could instead be used in this small-$\beta$ case, however in our proof of principle example, we keep the displacement duration fixed such that the pulses do not occupy a larger bandwidth than those used in the calibration.

The form of the resulting conditional displacement strongly depends on $\beta$ and the choice of $\alpha_0$. In \cref{fig:CD drives b} and \cref{fig:CD drives c} we illustrate the result of this optimization using our system parameters for $\beta = 1$ in two different regimes, $\alpha_0 = 10$ and $\alpha_0 = 50$. In the first case, a majority of the conditional displacement is accumulated during the wait times $t_w$. In the second case, $t_w$ is reduced to $0$, resulting in $\alpha_0$ being further lowered to $\approx 45$ to realize the gate. In this regime, the conditional displacement is accumulated during the driving periods, and increasing the target $\alpha_0$ does not result in a faster gate after optimization. This is the reason for the drive-constraint limit shown in Fig. 2c of the main text. 

\begin{figure*}[!ht]
	\begin{center}
		\includegraphics{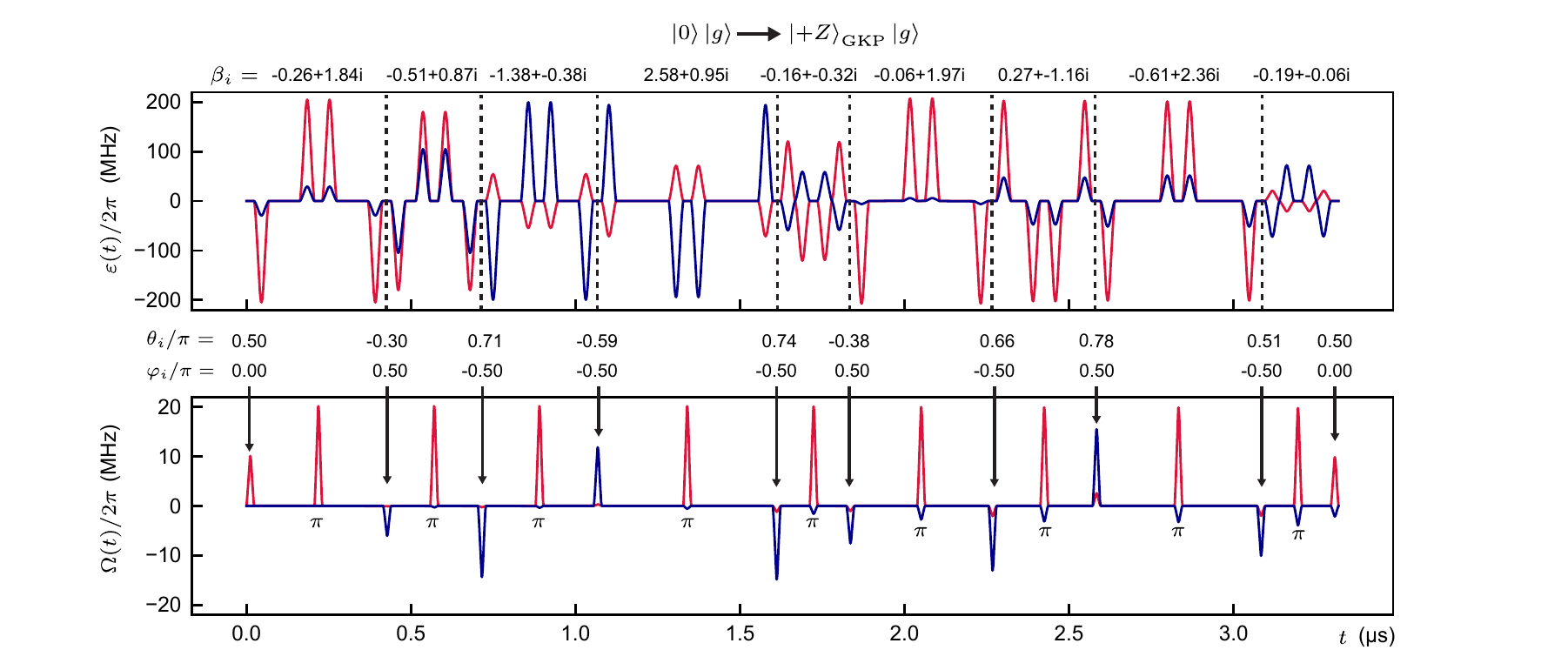}
		\caption{\label{fig:GKP pulse} \textbf{GKP state preparation ECD pulse sequence} used to prepare $\ket{+Z}_\text{GKP}$ in the cavity starting from vacuum using a circuit depth $N=9$ (here, $\Delta_\text{target} = 0.306$ as described in \cref{sec:GKP state analysis} and the ECD control parameters $\left\{\vec\beta,\vec\phi, \vec\theta \right\}$ are found using the procedure described in \cref{sec:ECD optimization}). $\varepsilon(t)$ is the cavity drive and $\Omega(t)$ is the transmon drive. Red and blue colors denote the real and imaginary parts of these drives, respectively.}
	\end{center}
\end{figure*}

A full ECD sequence specified by the parameters $\left\{\vec\beta,\vec\phi, \vec\theta \right\}$ is compiled into drives $\varepsilon(t)$ and $\Omega(t)$ by first optimizing the drives for each ECD gate given a target $\alpha_0$. These are interleaved with qubit rotations, which are performed by modifying the phase and amplitude of fixed-length Gaussian pulses with $\sigma = \SI{6}{ns}$ and total length $4\sigma = \SI{24}{ns}$ independently calibrated in experiment. We note that pulse shaping techniques such as derivative reduction by adiabatic gate (DRAG) could also be incorporated to realize shorter transmon pulses \cite{motzoiSimplePulsesElimination2009,chenMeasuringSuppressingQuantum2016a}. Finally, the phase of each qubit pulse is updated to account for the additional phases associated with each ECD gate ($\theta'$) calculated by \cref{eq:sol start}. This correction is done by keeping track of the qubit frame given all preceding ECD gates and updating the phase $\varphi$ of each qubit rotation gate accordingly. As an example, in \cref{fig:GKP pulse} we show the compiled ECD pulse sequence used to prepare $\ket{+Z}_\text{GKP}$ in the cavity. Here, we use $\alpha_0 = 30$, however some ECD gates are performed at a smaller $\alpha_0' < \alpha_0$ resulting from the finite displacement pulse duration and the ECD optimization procedure described above.

Finally, we note that the ECD gate set is designed to be well suited in the weak-$\chi$ regime since it requires fast unselective qubit rotations, an important operation that can become challenging at large dispersive shifts. With independent experiments on a different sample not presented in this work, we have confirmed the validity of the ECD approach for $\chi/2\pi \sim \SI{200}{kHz}$. To realize faster gates using a larger dispersive shift on the order of $\chi/2\pi \gtrsim \SI{1}{MHz}$, the gate set could be modified to take the partially selective nature of the finite-bandwidth qubit rotations into account, or GRAPE based techniques could be incorporated. 

\section{Characteristic function tomography}

\label{sec:tomography}
\subsection{Measurement and Post-Processing}

\begin{figure*}[h]
	\begin{center}
		\vspace{-2\baselineskip}
		\phantomsubfloat{\label{fig:tomography a}}
		\phantomsubfloat{\label{fig:tomography b}}
		\phantomsubfloat{\label{fig:tomography c}}
		\includegraphics{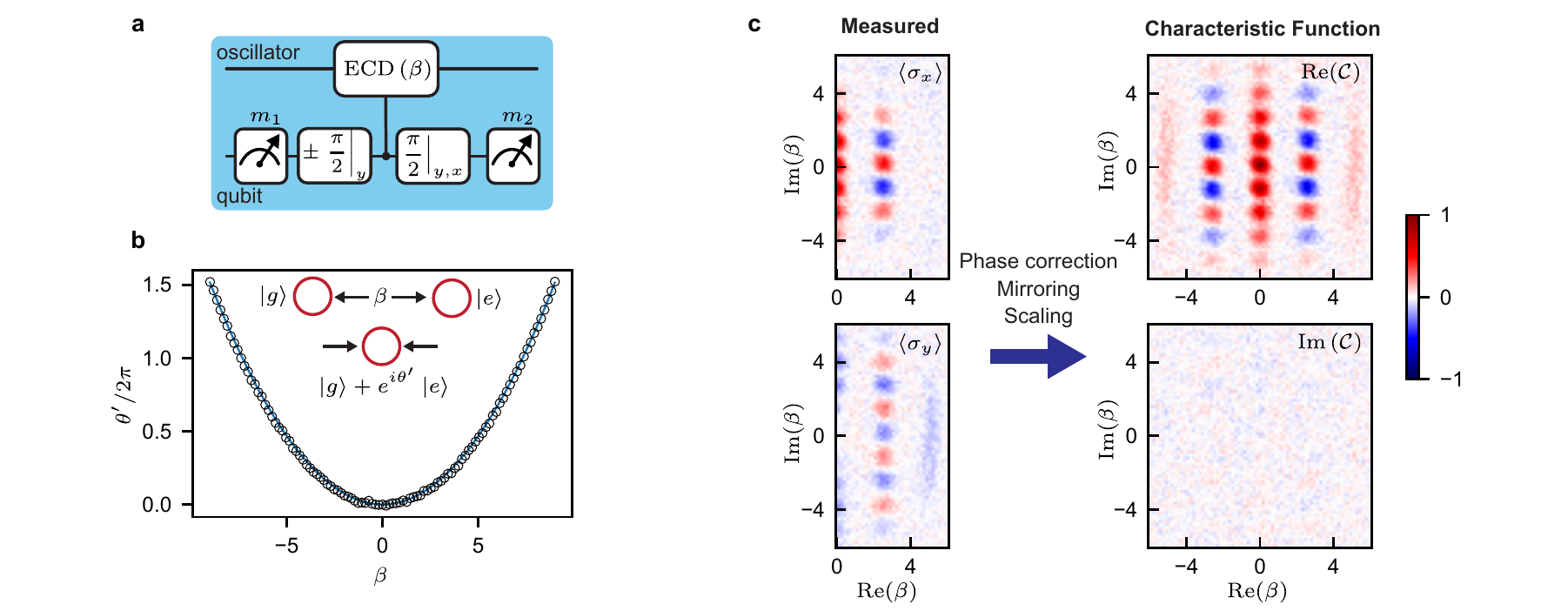}
		\caption{\label{fig:tomography}\textbf{Tomography} \textbf{a.} Characteristic function measurement sequence \textbf{b.} Measured qubit phase accumulation (open circles) after the \textit{cat-and-back} sequence (depicted in the inset). Data matches the phase predicted by \cref{eq:sol end} (solid line). \textbf{c.} Example of characteristic function post-processing using the measured GKP $\ket{+Z}$ state.}
	\end{center}
\end{figure*}

The tomographic sequence used to measure the characteristic function after each ECD sequence is shown in \cref{fig:tomography a}. As shown in previous works (\cite{fluhmannDirectCharacteristicFunctionTomography2020, fluhmannEncodingQubitTrappedion2019, campagne-ibarcqQuantumErrorCorrection2020}), the oscillator's characteristic function defined as $\mathcal{C}(\beta) = \text{Tr}\left(D(\beta)\rho\right)$ can be measured using a conditional displacement embedded within a qubit ramsey sequence resulting in $\braket{\sigma_x - i\sigma_y} = \braket{D(\beta)}$ before the second $\pi/2$ pulse. By varying the phase of the second $\pi/2$ pulse, we can measure either the real or imaginary part of $\mathcal{C}(\beta)$ by measuring either $\braket{\sigma_x}$ or $\braket{\sigma_y}$ respectively. We also include a first measurement $m_1$ to disentangle the qubit and oscillator before the tomography, and postselect the results of the characteristic function ($m_2$) on $m_1 = \ket{g}$. We note that the pulses in this work are designed to realize state preparation sequences of the form $\ket{0}\ket{g} \rightarrow \ket{\psi}\ket{g}$ which disentangle the oscillator and qubit after the pulse. However, due to decoherence during the pulse, there is small residual entanglement between the oscillator and qubit, hence the need for $m_1$, with probabilities of $m_1 = \ket{g}$ given in the main text. 

To simplify tomography, the ECD gate used to measure the characteristic function is realized using the optimization method in \cref{sec:ECD optimization} for $\beta=1.0$ using $\alpha_0=2.0$. The amplitude and phase of this ECD gate is swept in experiment to sample the characteristic function. This low $\alpha_0$ is chosen so the second-order dispersive shift and Kerr have a negligible effect on the tomography, and the applied gate is well described by \cref{eq:cd_unitary}. From this, the additional qubit phase $\theta'$ can be easily predicted and corrected in post-processing.

To verify the validity of \cref{eq:sol start} in predicting this phase, we measure the phase $\theta'$ by using a \textit{cat-and-back} ECD sequence depicted in the inset of \cref{fig:tomography b}: We prepare the qubit in $\ket{+x} = \frac{1}{\sqrt{2}}\left(\ket{g} + \ket{e}\right)$, then apply two conditional displacements: $\text{ECD}(\beta), \text{ECD}(-\beta)$, with a qubit $\pi$ pulse between, after which the qubit phase is measured (by measuring $\braket{\sigma_x}$ and $\braket{\sigma_y}$). Here, the $\text{ECD}$ gates are the same as those used for characteristic function tomography ($\alpha_0 = 2$ at $\beta=1$, and amplitude is swept). The resulting phase $\theta'$ in \cref{fig:tomography b} shows excellent agreement with the prediction from \cref{eq:sol start}.

With this, the post-processing of the tomographic $\braket{\sigma_x}$ and $\braket{\sigma_y}$ is depicted in \cref{fig:tomography c}, using the measured data for the $\ket{+Z}_\text{GKP}$ state as an example. In experiment, we alternate between $\pm \pi/2$ for the first $\pi/2$ pulse to symmetrize the transmon's $T_1$ error channel during readout. Since the characteristic function obeys the property $\mathcal{C} (-\beta) = \mathcal{C}^*(\beta)$, we only measure half of the real and imaginary parts, then mirror about the $\text{Re}(\beta) = 0$ axis. The characteristic function is found by applying a phase correction $\mathcal{C}(\beta) \rightarrow e^{i|\beta|^2\theta'_0}\mathcal{C}(\beta)$ where $\theta'_0$ is the phase predicted by \cref{eq:sol start} for $\beta = 1$ using the ECD pulse optimized for tomography ($\alpha_0 = 2$). We note that this phase is slightly different than the phase found using the cat-and-back experiment described above, which is only verifying the validity of \cref{eq:sol start}. Finally, the data is scaled such that $\mathcal{C}(0) = 1$, effectively accounting for qubit decoherence ($T_{2E}$) during the tomography.

\subsection{Density matrix reconstruction}

\label{sec:density matrix recon}
To estimate the fidelity and purity of the oscillator state in experiment, we employ density matrix reconstruction using maximum likelihood estimation. For this, we use the measured real and imaginary parts of the characteristic functions, taken with 1,280 averages per point and use a numerical, iterative, convex optimization algorithm. For all demonstrations, the measured imaginary parts are close to zero, as is expected for states with symmetric Wigner functions. Any small deviations away from zero in the imaginary part are captured by the reconstruction. 

For the Fock states, binomial states, and GKP states, the reconstruction is done in the Fock basis. For the squeezed states, reconstruction is performed in the squeezed-Fock basis with a basis squeezing equal to the target squeezing. The reconstruction Hilbert space size is swept, and a Hilbert space trunction is chosen such that increasing or decreasing the truncation does not change the fidelity or purity within the quoted error bars. For the binomial,GKP, and squeezed states, some states display a small phase-space rotation in the tomography. For these, a small inverse rotation is applied to the reconstructed density matrix. The maximum change in fidelity from this rotation is $1(\%)$ for the $\ket{-Z}_\text{GKP}$ state. 

\subsection{Effective squeezing measurement}

\begin{figure*}[!h]
	\begin{center}
		\vspace{-2\baselineskip}
		\phantomsubfloat{\label{fig:squeezing a}}
		\phantomsubfloat{\label{fig:squeezing b}}
		\includegraphics{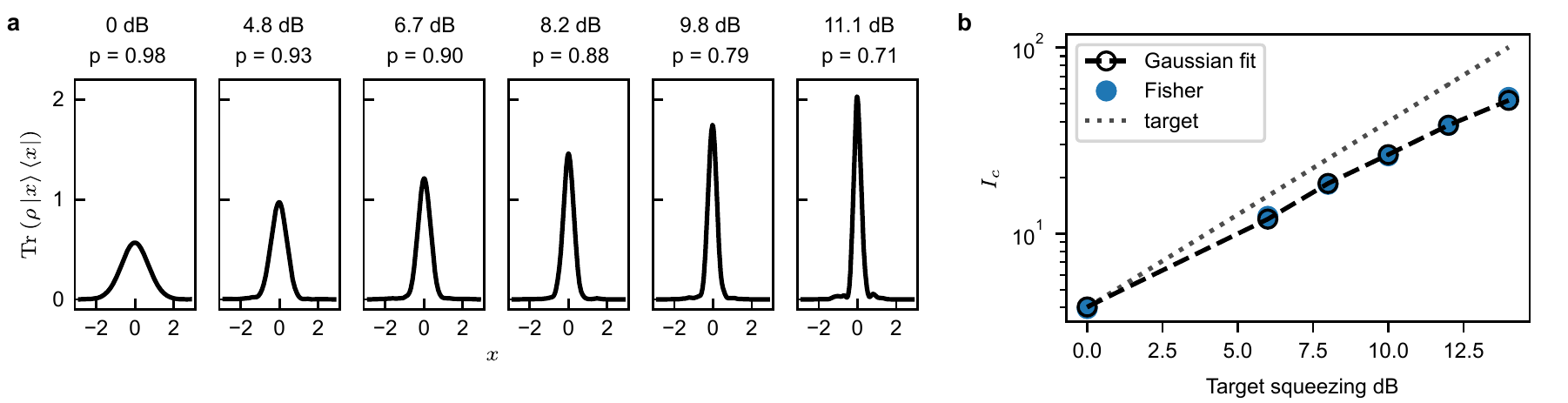}
		\caption{\label{fig:squeezing} \textbf{Squeezed state distributions} \textbf{a.} Reconstructed position quadrature probability distributions and squeezing values found by a fit to Gaussian functions. Also shown are purities calculated as $p = \text{Tr} (\rho^2)$. \textbf{b.} Calculated Fisher information from the probability distributions (orange), compared to that expected from a Gaussian fit (blue).}
	\end{center}
\end{figure*}

To find the effective squeezing of prepared states, the reconstructed density matrices are used to calculate the position quadrature probability distributions, $P(x) = \text{Tr}\left(\rho \ket{x}\bra{x}\right)$, with results shown in \cref{fig:squeezing a}, including the measured cavity equilibrium thermal state for comparison. Here, a small rotation is applied to the reconstructed density matrices before calculating the probability distributions to align the squeezed quadrature. Also shown is the purity ($p = \text{Tr}\left(\rho^2\right)$) of each reconstructed density matrix. These distributions are fit to Gaussian functions to extract the squeezing in each state quoted in the main text. We compare these results to a calculation of the Fisher information using the full reconstructed probability distributions,
\begin{align}
	I_c = 2\int dx\, \left(\partial_x \log P(x)\right)^2 P(x).
\end{align}
The Fisher information is a measure of the ability to sense small displacements along the position quadrature using the state with respect to homodyne detection \cite{parisQuantumStateEstimation2004, hastrupUnconditionalPreparationSqueezed2021}. For ideal squeezed states, $I_{c} = 2/\braket{\Delta x^2}$. Although the calculated purity of the squeezed states decreases with larger target squeezing, the states can still be used to sense small displacements, since only the probability distribution $P(x) = \text{Tr}\left(\rho \ket{x}\bra{x}\right)$ enters into the Fisher information. In \cref{fig:squeezing b}, we compare the calculated $I_C$ for each state to the target value. Finally, we also compare to $I_c = 2/\sigma_x^2$, where $\sigma_x^2$ is the variance of the Gaussian fit, indicating that the squeezed quadrature distributions are well approximated by a Gaussian, even though the full tomographies show some non-Gaussian features.

\subsection{Binomial code analysis}

The binomial \textit{kitten code} is the smallest binomial code for which the loss of a single photon is correctable \cite{michaelNewClassQuantum2016,huQuantumErrorCorrection2019, grimsmoQuantumComputingRotationsymmetric2020, gertlerProtectingBosonicQubit2021,burkhartErrorDetectedStateTransfer2021}, with logical states given by $\ket{+Z} = \left(\ket{0} + \ket{4}\right) / \sqrt{2}$ and  $\ket{-Z} = \ket{2}$. In \cref{tab:Binomial expectation values} we give the estimated expectation values of the logical Pauli operators of the prepared states found numerically using the reconstructed density matrices. We also quantify how errors in the prepared states could be corrected by ideal error correction, as some errors during the state preparation are in principle correctable. In particular, the correctable code space includes the normalized error states $\ket{+Z}_\text{e} = a\ket{+Z}/ |\braket{+Z|a^\dagger a |+Z}|^2 = \ket{3}$ and  $\ket{-Z}_\text{e} = a\ket{-Z}/ |\braket{-Z|a^\dagger a |-Z}|^2 = \ket{1}$. From these, we can define the logical operators corresponding to applying ideal error correction then performing a logical measurement, given by
\begin{equation}
	\begin{aligned}
		&X_c = \ket{+Z}\bra{-Z} + \ket{+Z}_\text{e}\bra{-Z}_\text{e} + \ket{-Z}\bra{+Z} + \ket{-Z}_e\bra{+Z}_e \\
		&Y_c = i\ket{-Z}\bra{+Z} + i\ket{-Z}_e\bra{+Z}_e - i\ket{+Z}\bra{-Z} - i\ket{+Z}_e\bra{-Z}_e\\
		&Z_c = \ket{+Z}\bra{+Z} + \ket{+Z}_\text{e}\bra{+Z}_\text{e} - \ket{-Z}\bra{-Z} - \ket{-Z}_e\bra{-Z}_e \\
		&I_c = \ket{+Z}\bra{+Z} + \ket{+Z}_\text{e}\bra{+Z}_\text{e} + \ket{-Z}\bra{-Z} + \ket{-Z}_e\bra{-Z}_e.
	\end{aligned}
\end{equation}
The expectation values of these operators quantify the logical information encoded in the prepared states after ideal error correction assuming a photon loss error channel. We calculate these expectation values using the reconstructed density matrices with results shown in \cref{tab:Binomial expectation values}. If these prepared states were to be used in an error correction setting, the error decoding model should instead be adapted to fit the actual errors encountered during state preparation. 

\begin{table}[!h]
	\begin{tabular}{c|cccc|cccc}
		State& $\braket{I}$  & $\braket{X}$ & $\braket{Y} $ & $\braket{Z}$ & $\braket{I_c}$ & $\braket{X_c}$ & $\braket{Y_c}$ & $\braket{Z_c}$\\
		\hline 
		$\ket{+X}_\text{bin}$ & 0.90 & 0.98 & 0.08 & -0.05 & 0.95 & 0.92 & 0.09 & -0.08 \\
		$\ket{+Y}_\text{bin}$ & 0.92 & -0.06 & 0.90 & -0.10 &  0.98 & -0.06 & 0.96 & -0.13 \\
		$\ket{+Z}_\text{bin}$ & 0.92 & 0.03 & -0.05 & 0.92 & 0.97 & 0.04 & -0.04 & 0.96\\
		$\ket{-Z}_\text{bin}$ & 0.93 & -0.01 & -0.10 & -0.93 & 0.99 & -0.02 & -0.10 & -0.98\\
	\end{tabular}
	\caption{\label{tab:Binomial expectation values} \textbf{Binomial code Pauli expectation values} found numerically using the reconstructed density matrices.}
\end{table}

\subsection{GKP code analysis}

\label{sec:GKP state analysis}
The finite energy square GKP code stabilizers and logical Pauli operators are defined as \cite{gottesmanEncodingQubitOscillator2001,menicucciFaultTolerantMeasurementBasedQuantum2014,nohQuantumCapacityBounds2019,royerStabilizationFiniteEnergyGottesmanKitaevPreskill2020,terhalScalableBosonicQuantum2020,grimsmoQuantumErrorCorrection2021}
\begin{align}
	&S_{q,\Delta} = e^{-\Delta^2 a^\dagger a} D\left(i\sqrt{2\pi}\right) e^{\Delta^2 a^\dagger a} &S_{p,\Delta} = e^{-\Delta^2 a^\dagger a} D\left(\sqrt{2\pi}\right) e^{\Delta^2 a^\dagger a} \\
	&X_{\Delta} = e^{-\Delta^2 a^\dagger a} D\left(\sqrt{\pi/2}\right) e^{\Delta^2 a^\dagger a} &Z_{\Delta} = e^{-\Delta^2 a^\dagger a} D\left(i\sqrt{\pi/2}\right) e^{\Delta^2 a^\dagger a}
\end{align}

with $Y_\Delta = iX_\Delta Z_\Delta $. The target GKP states are found numerically by letting $\ket{+Z_\Delta}$ be the groundstate of a fictitious Hamiltonian $H = -S_{q,\Delta} - S_{p,\Delta} - Z_\Delta$, then by applying the appropriate finite energy logical operators and normalizing. Here, we use a target state squeezing of $\Delta_\text{target} = 0.306$ (10.3 dB). To estimate the effective squeezing of the prepared GKP states in experiment, we find the value of $\Delta$ that maximizes the expectation value of the projector onto the finite-energy code space $P_\Delta = \ket{+Z_\Delta}\bra{+Z_\Delta} + \ket{-Z_\Delta}\bra{-Z_\Delta}$ using the reconstructed density matrices. We find a squeezing of $\Delta_\text{exp} = 0.35$ (9.1 dB) for all prepared states within the precision of the reconstruction. In addition to the fidelities given in the main text, we quantify the quality of the prepared states here by numerically estimating the expectation values of the finite-energy Pauli operators and stabilizers with results given in \cref{tab:GKP expectation values}.

\begin{figure*}[ht]
	\begin{center}
		\includegraphics{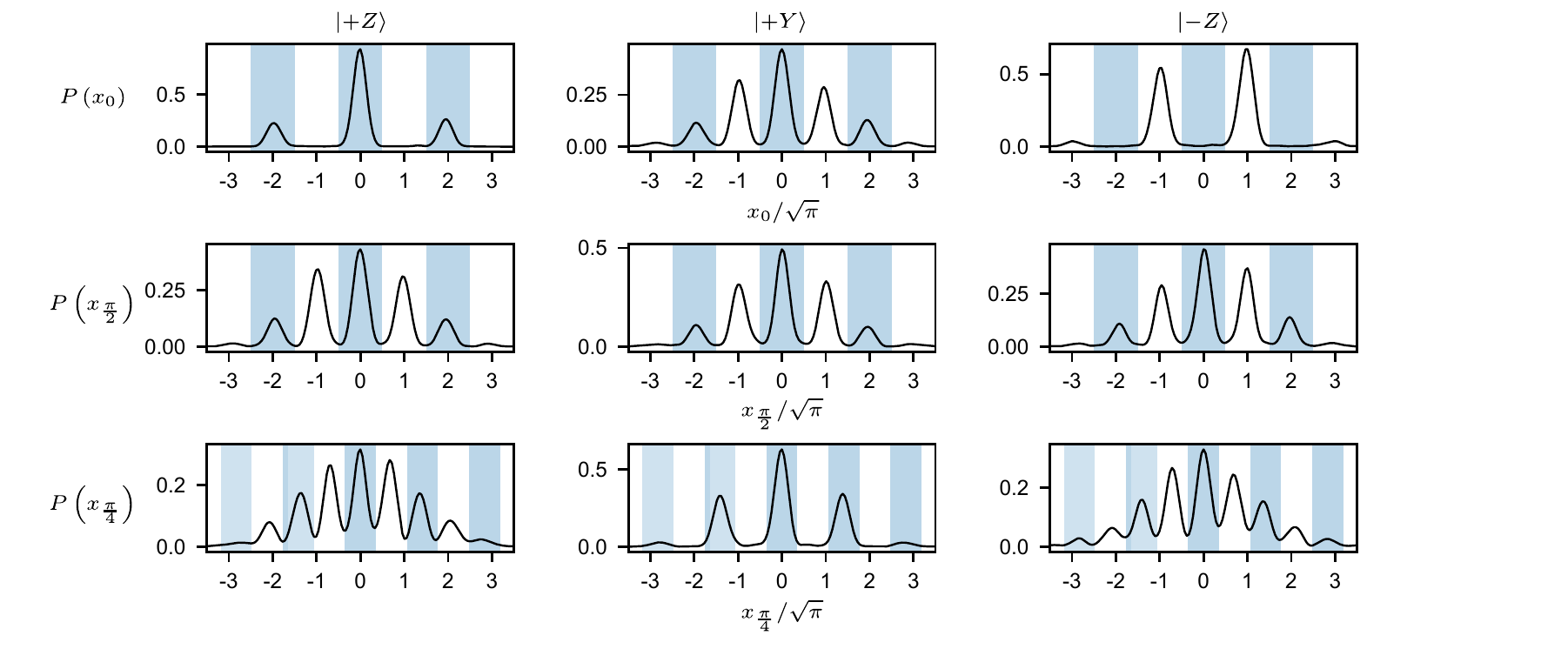}
		\caption{\label{fig:GKP marginals} \textbf{Reconstructed GKP state marginal probability distributions.} Blue and white bins represent the integration zones corresponding to assigning the associated logical Pauli operator $\pm 1$ respectively.}
	\end{center}
\end{figure*}

In addition, we quantify the prepared GKP states by the logical expectation values that would result from an ideal homodyne detection. In \cref{fig:GKP marginals}, we plot the reconstructed marginal probability distributions for the prepared GKP states along the generalized quadrature coordinate $x_\theta = \left(e^{i\theta }a^\dagger + e^{-i\theta}a\right)/\sqrt{2}$ with $\theta = \left\{0,\pi/2, \pi/4\right\}$ ($x_0 = x$ and $x_{\frac{\pi}{2}} = p$). From these, we can define the corresponding homodyne expectation values ($X_H$, $Y_H$, and $Z_H$) resulting from integrating the probability distributions and assigning a logical value. These are given by the total probability of finding a homodyne measurement result in sectors closest to $x_0 \mod 2\sqrt{\pi} = 0$  or $x_0 \mod 2\sqrt{\pi} = \sqrt{\pi}$ for $Z_H = \pm 1$, with analogous definitions for $X_H$ and $Y_H$ \cite{gottesmanEncodingQubitOscillator2001}. The results are given in \cref{tab:GKP expectation values}.

\begin{table}[!h]
	\begin{tabular}{c|ccccc|ccc}
		State& $\text{Re}\left(\braket{S_{q,\Delta}}\right)$  & $\text{Re}\left(\braket{S_{p,\Delta}}\right)$  & $\text{Re}\left(\braket{X_\Delta}\right)$ & $\text{Re}\left(\braket{Y_\Delta}\right) $ & $\text{Re}\left(\braket{Z_\Delta}\right)$ & $\braket{X_\text{H}}$ & $\braket{Y_\text{H}}$ & $\braket{Z_\text{H}}$\\
		\hline 
		$\ket{+Z}_\text{GKP}$ & 0.75     & 0.88 & 0.01 & -0.02 & 0.94 & 0.00 & -0.01 & 0.94\\
		$\ket{+Y}_\text{GKP}$ &  0.78     & 0.78 &0.02&0.87&0.05 &0.01 & 0.87 & 0.06 \\
		$\ket{-Z}_\text{GKP}$ & 0.81      & 0.71 &0.02 & 0.03 & -0.85 & 0.02 & 0.02 & -0.91  \\
	\end{tabular}
	\caption{\label{tab:GKP expectation values} \textbf{GKP code stabilizer and Pauli expectation values}. For the finite energy stabilizers and Pauli operators, we use $\Delta = \Delta_\text{exp} = 0.35$.}
\end{table}

\section{Sources of Infidelity}

In this section, we use simulations to estimate the sources of infidelity for the Fock state, binomial state, and GKP state preparation examples shown in the main text. 

\label{sec:error budget}
\subsection{Decoherence-Free Error Budget}

\begin{figure*}[!ht]
	\begin{center}
		\includegraphics{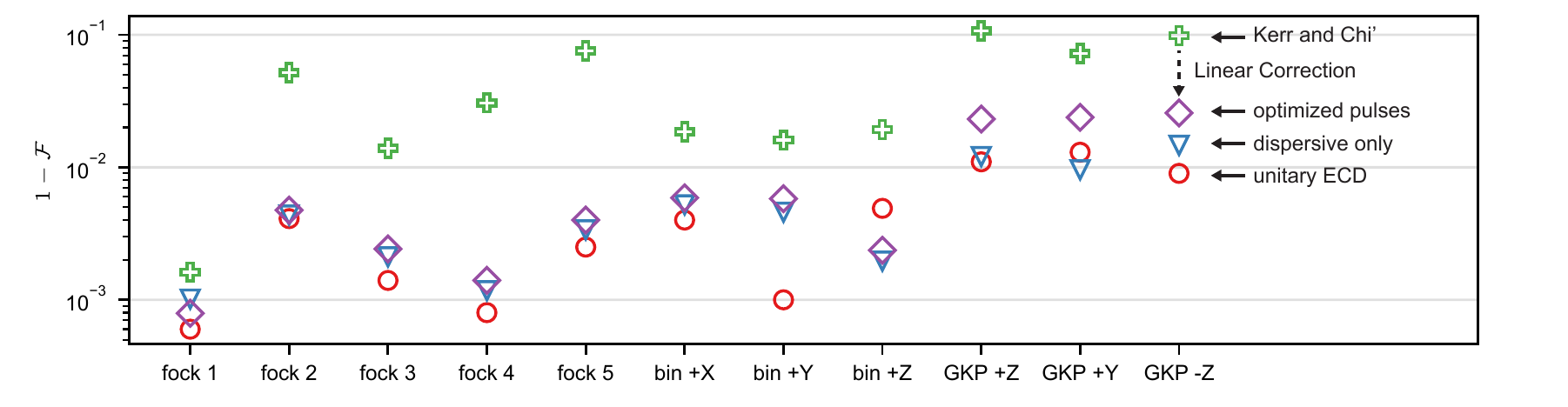}
		\caption{\label{fig:lossless error budget} \textbf{Decoherence free error budget} Simulated fidelity of constructed ECD sequences under different conditions. Red circles: result from unitary ECD parameter optimization. Blue triangles: pulses constructed and simulated with $K = \chi' = 0$. Green crosses: pulses optimized with $K = \chi' = 0$ and simulated with measured $K$ and $\chi'$ (nonzero). Purple diamonds: pulses optimized and simulated with measured $K$ and $\chi'$ (nonzero)f.}
	\end{center}
\end{figure*}

First, we estimate how accurately our pulse compilation procedure, described in \cref{sec:ECD optimization}, can realize ideal ECD control unitaries, $U_\text{ECD}$ in \cref{eq:ECD unitary}, especially in the presence of Kerr and the second-order dispersive shift. In the open red circles of \cref{fig:lossless error budget}, we show simulated oscillator state preparation fidelities of the ideal ECD unitaries, defined as $\mathcal{F}_g = |\braket{\psi_\text{target}|\psi_g}|^2$
where $\ket{\psi_g}$ is the oscillator's state after postselecting the qubit in $\ket{g}$.

Next, we use the pulse compilation procedure described in \cref{sec:ECD optimization} with our system parameters, except we set $K_s = 0$ and $\chi' = 0$ to construct oscillator and qubit pulses which realize the ECD sequences without these higher order nonlinearities. 
These pulses are simulated with $K_s = 0$ and $\chi' = 0$, and the resulting fidelities are shown by the blue triangles in \cref{fig:lossless error budget}.
These fidelites are close to the ideal unitary fidelities, demonstrating our ability to realize ECD sequences with the ideal dispersive Hamiltonian. These pulses are then simulated using \cref{eq:displaced H} with all nonlinear terms included using the measured system parameters, and the results are shown by the green crosses in \cref{fig:lossless error budget}. These infidelities are significantly higher, demonstrating the need to account for higher order nonlinaerities in the pulse construction. 

Finally, we include the measured values of $K_c$ and $\chi'$ in the pulse construction, following the procedure outlined in \cref{sec:ECD optimization} which corrects for the linear contributions of Kerr and the second-order dispersive shift in the displaced frame. The resulting pulses are simulated with the full Hamiltonian, resulting in the infidelities given by purple diamonds in \cref{fig:lossless error budget}. These infidelities are close to the pulses optimized and simulated with $K=0$ and $\chi'=0$, indicating that correcting for the linear contributions of these terms is enough to significantly reduce their deleterious impact on the overall gate fidelity.

\subsection{Impact of decoherence}

Next, we study the impact of transmon and cavity decoherence on the state preparation fidelity. In particular, we perform master equation simulations in the time-dependent displaced frame. Under a unitary frame transformation $\rho \rightarrow \tilde{\rho} = U\rho U^\dagger$, the master equation $\partial_t \rho = -(i/\hbar) \left[H,\rho\right] + \sum_i \mathcal{D}[L_i]\rho$ becomes $\partial_t \tilde{\rho} =  -(i/\hbar) \left[\tilde{H},\rho\right] + \sum_i \mathcal{D}[U L_i U^\dagger]\tilde{\rho}$ with $\tilde{H} = UH U^\dagger + i\hbar \left(\partial_t U\right)$. Using the time-dependent displaced frame in \cref{sec:trajectories}, we evolve the joint transmon-cavity density matrix according to
\begin{equation}
	\begin{aligned}
		\label{eq:master equation}
		&\partial_t \tilde{\rho} = -\frac{i}{\hbar}\left[\tilde{H}(t), \tilde{\rho}\right] + \gamma_\downarrow \mathcal{D}[q]\tilde{\rho} + \gamma_\uparrow \mathcal{D}[q^\dagger] \tilde{\rho} + 2\gamma_\phi \mathcal{D}[q^\dagger q]\tilde{\rho} \\
		&+  \kappa_\downarrow \mathcal{D}[a]\tilde{\rho} + \kappa_\uparrow \mathcal{D}[a^\dagger + \alpha^*(t)]\tilde{\rho} + 2 \kappa_\phi \mathcal{D}[(a^\dagger + \alpha^*(t))(a + \alpha(t))]\tilde{\rho}
	\end{aligned}
\end{equation}
where $\tilde{H}(t)$ is the displaced frame Hamiltonian in \cref{eq:displaced H} and $\alpha(t)$ is the nonlinear response to the drive given by solving \cref{eq:cm trajectory}. By simulating in the displaced frame which tracks the classical trajectory of the state's center-of-probability in phase space, we reduce the Hilbert space truncation required to accurately capture the dynamics. This is especially important considering our pulses drive the oscillator to photon numbers of $\sim 10^3$.

\begin{figure*}[!ht]
	\begin{center}
		\includegraphics{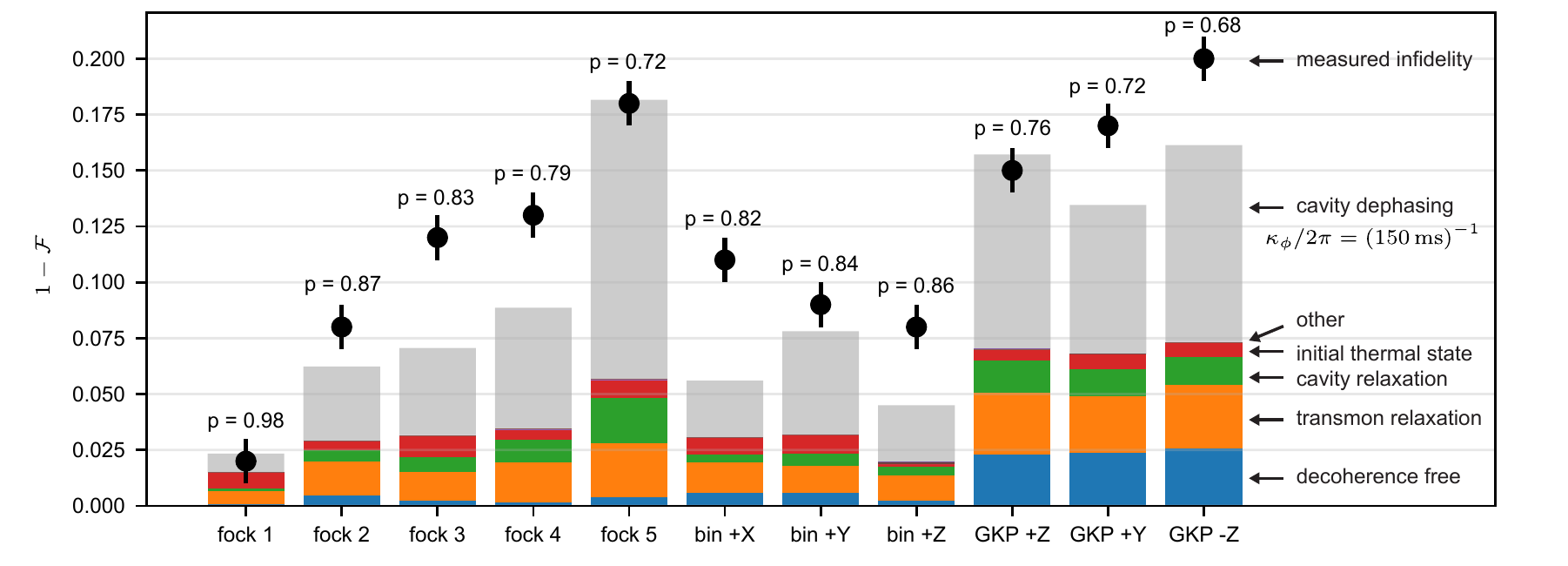}
		\caption{\label{fig:error budget} \textbf{Measured fidelities, purities, and simulated error budget} For each state preparation example, black dots indicate infidelities found from maximum likelihood reconstruction using the measured characteristic functions after postselecting the transmon in $\ket{g}$. Also shown is the purity of the reconstructed states $p = \text{Tr}\left(\rho^2\right)$. For each state, we include the baseline infidelity (labeled \textit{decoherence-free}) which is equivalent to the purple diamonds in \cref{fig:lossless error budget}. On top of this baseline, contributions to the total simulated infidelity using the measured decoherence rates are shown by the colored bars. Here, \textit{other} (bars not visible) includes contributions from transmon dephasing, transmon heating, and cavity heating at the quoted rates. Also included is simulated contribution due to intrinsic cavity dephasing at a rate $\kappa_\phi/2\pi = \left(\SI{150}{ms}\right)^{-1} \approx \SI{1}{Hz}$.}
	\end{center}
\end{figure*}

We first simulate the impact of transmon relaxation, heating, and dephasing, as well as cavity relaxation and heating, given the measured rates in \cref{tab:system parameters}. When simulating transmon relaxation, we use the averaged measured relaxation rate with $\braket{n}_\text{cav} = 900$ photons in the cavity $\tilde{T}_{1,q}\approx \SI{30}{us}$, however we note this changes with time as shown in \cref{fig:out and back}, and the contribution from this error channel is also expected to change depending on the particular $\tilde{T}_{1,q}$ at the time of averaging. In addition, we approximate the qubit dephasing rate to be $\gamma_\phi = \gamma_{2E} - \gamma_1/2$, which assumes the echoing of low-frequency noise during the ECD gates results in an effective white noise dephasing.  

The results are summarized by the colored bars in \cref{fig:error budget}, where each contribution to the error budget is simulated by only including a single decoherence mechanism. Also included is infidelity associated to an initial cavity thermal state ($n_\text{th} = 0.025$) since we do not employ active cavity cooling before each experiment. Independent simulations have verified that adding the infidelities of individual error channels is a good predictor of the total infidelity when simulating with all error channels combined. Out of these error channels, transmon relaxation has the biggest overall impact on the infidelity as ancilla relaxation during the conditional displacements can result in large oscillator displacements \cite{campagne-ibarcqQuantumErrorCorrection2020}.

\subsection{Discussion}

As shown in \cref{fig:error budget}, these decoherence mechanisms alone under-predict the infidelity found in experiment. Possible additional sources of infidelity include unknown transfer functions \cite{gustavssonImprovingQuantumGate2013,rolRestlessTuneupHighFidelity2017a,jergerSituCharacterizationQubit2019}, drifts in parameters, additional cavity heating due to the strong drives, and cavity dephasing. Out of these mechanisms, we simulate the effect of cavity dephasing using the displaced-frame dephasing term in \cref{eq:master equation}. 

In this experiment, we do not have a direct measurement of the small intrinsic cavity dephasing  rate, and such a measurement is an ongoing topic of investigation. We instead use master equation simulations to study the impact of cavity dephasing. A previous work on the same physical sample has also used simulations to bound the dephasing rate to $\kappa_\phi \lesssim \SI{1}{Hz}$ \cite{campagne-ibarcqQuantumErrorCorrection2020} and here we find a similar result. In \cref{fig:error budget} we include a contribution to the infidelity by simulating the pulses with a cavity dephasing rate of $\kappa_\phi/2\pi = \left(\SI{150}{ms}\right)^{-1} \approx \SI{1}{Hz}$, shown by the light grey bars. For some state preparation experiments, such as Fock $\ket{5}$ and GKP $\ket{+Z}$, this rate roughly matches the measured infidelity. This is evidence that the dephasing rate is likely smaller than $\SI{1}{Hz}$, matching the result from \cite{campagne-ibarcqQuantumErrorCorrection2020}. However, for other experiments, the error bars still under predict the fidelity, indicating other unknown mechanisms.

\section{Optimization of ECD circuit parameters}

\label{sec:numerical optimization}
For state preparation using ECD control, the quantum control problem we aim to solve is
\begin{align}
	\label{eq:ECD unitary}
	&U_\text{ECD} = D(\beta_{N+1}/2)R_{\varphi_{N+1}}\left(\theta_{N+1}\right)\text{ECD}(\beta_N)R_{\varphi_N}\left(\theta_N\right)...\text{ECD}(\beta_2)R_{\varphi_2}\left(\theta_2\right)\text{ECD}(\beta_1)R_{\varphi_1}\left(\theta_1\right) \\
	\label{eq:full fidelty}
	&\mathcal{F} = |\braket{\psi_t|U_\text{ECD}|\psi_i}|^2\\
	&\left\{\vec{\beta}, \vec{\varphi}, \vec{\theta} \right\} = \argmax_{\left\{\vec{\beta}, \vec{\varphi}, \vec{\theta} \right\}} (\mathcal{F})
\end{align}
with initial state $\ket{\psi_i}$ and target state $\ket{\psi_t}$.
The circuit depth $N$ should be chosen such that $\mathcal{F}$ at its maximum is above an acceptable value with experimental considerations in mind. Although we focus on state preparation in this work, the optimization method described can also be used to realize a general unitary $U_\text{target}$ by replacing the Fidelity function with $\mathcal{F} = \left| \frac{1}{\text{Tr}(P)} \text{Tr}\left(PU_\text{target}^\dag U_\text{ECD}\right)\right|^2$, where $P$ is a projector onto a possible subspace of interest \cite{ballSoftwareToolsQuantum2021}. In all protocols presented in this work we include a final qubit rotation $R_{\varphi_{N+1}}\left(\theta_{N+1}\right)$ and displacement $D(\beta_{N+1}/2)$ after the last ECD gate in the optimizer. Often, the optimizer converges to protocols with $\beta_{N+1} = 0$. These gates are implemented quickly with respect to typical ECD gates and are not included in the quoted circuit depths $N$ which only counts the total number of ECD gates.

Here, we realize a multi-start method to solve this non-convex problem by optimizing $B$ independent variations of $U_\text{ECD}$ in a parallel manner.
Denoting the $j^{\text{th}}$ variation of the ECD unitary as $U_{\text{ECD},j}$ with circuit parameters $\left\{\vec{\beta}_j, \vec{\varphi}_j, \vec{\theta}_j \right\}$ and fidelity $\mathcal{F}_j = |\braket{\psi_t|U_{\text{ECD},j}|\psi_i}|^2$, we perform gradient descent on all $4BN$ real parameters using the total cost function $\text{C} = \sum_{j = 1}^B \log ( 1 - \mathcal{F}_j)$. Since the cost function is a simple sum of independent logarithmic cost functions, gradient descent of $\text{C}$ realizes independent gradient descent of each circuit realization in parallel. The optimization is stopped when any $\mathcal{F}_j$ reaches the target fidelity, and the parameters from the $j^\text{th}$ circuit are selected.

We realize gradient descent of the cost function using \textit{Adam} \cite{kingmaAdamMethodStochastic2017} implemented in TensorFlow. To construct this cost function and its gradient, we represent the batch of circuit parameters by the tensors $\bm{\beta}$, $\bm{\varphi}$ and $\bm{\theta}$ of dimensions $B \times N$ such that $\bm{\beta}_{ji}$ is the $i^\text{th}$ parameter appearing in circuit $j$.
Given a tensor-product structure of $\mathcal{H} = \mathbb{C}^2 \otimes \mathbb{C}^{N_o}$ where $N_o$ is the truncation of the oscillator's Hilbert space, the ECD unitaries and cost function are represented in the block-matrix form 
\begin{align}
	&U_{\text{ECD}, j} = \bm{b}_{jN}...\bm{b}_{j2}\bm{b}_{j1} \\
	&\bm{b}_{ji} = \text{ECD}(\bm{\beta}_{ji})R_{\bm{\varphi}_{ji}}(\bm{\theta}_{ji}) = 
	\begin{pmatrix}
		D^\dag(\bm{B}_{ji})e^{i\bm{\Phi}_{ji}}\sin\bm{\Theta}_{ji} & 	 D^\dag(\bm{B}_{ji})\cos\bm{\Theta}_{ji} \\
		D(\bm{B}_{ji})\cos\bm{\Theta}_{ji} &  -D(\bm{B}_{ji})e^{-i\bm{\Phi}_{ji}}\sin\bm{\Theta}_{ji}
	\end{pmatrix} \\
	&\text{C} = \sum_{j=1}^{N} \log\left(1 -|\bra{\psi_t} \bm{b}_{jN}...\bm{b}_{j2}\bm{b}_{j1}\ket{\psi_i}|^2\right)
\end{align}
with reduced parameters $\bm{B} = \frac{\bm{\beta}}{2}$, $\bm{\Theta} = \frac{\bm{\theta}}{2}$ and $\bm{\Phi} = \bm{\varphi} - \frac{\pi}{2}$, and $D$ is the displacement operator defined on the oscillator's Hilbert space truncated to $N_o$. To construct each block operator $\bm{b}_{ji}$, we first compute the displacement operators. For this, we use the batched displacement operator implementation in \cite{sivakModelFreeQuantumControl2021}, which uses pre-diagonalization of the truncated position and momentum operators to efficiently construct the displacement matrices. With this, we compute all $B\times N$ displacement opertors $D(\bm{B})$ in parallel, then build the block matrices $b_{ji}$ block-by-block while reusing computed functions to minimize the total number of computations. Once these blocks are computed, the cost function is implemented by contracting along the $i$ index, taking the logarithm, then contracting along the $j$ index. To compute the gradient of the cost function with respect to $\bm{\beta}$, $\bm{\varphi}$, and $\bm{\theta}$, we use TensorFlow to realize reverse-accumulation automatic differentiation. 

\begin{figure*}[!ht]
	\begin{center}
		\includegraphics{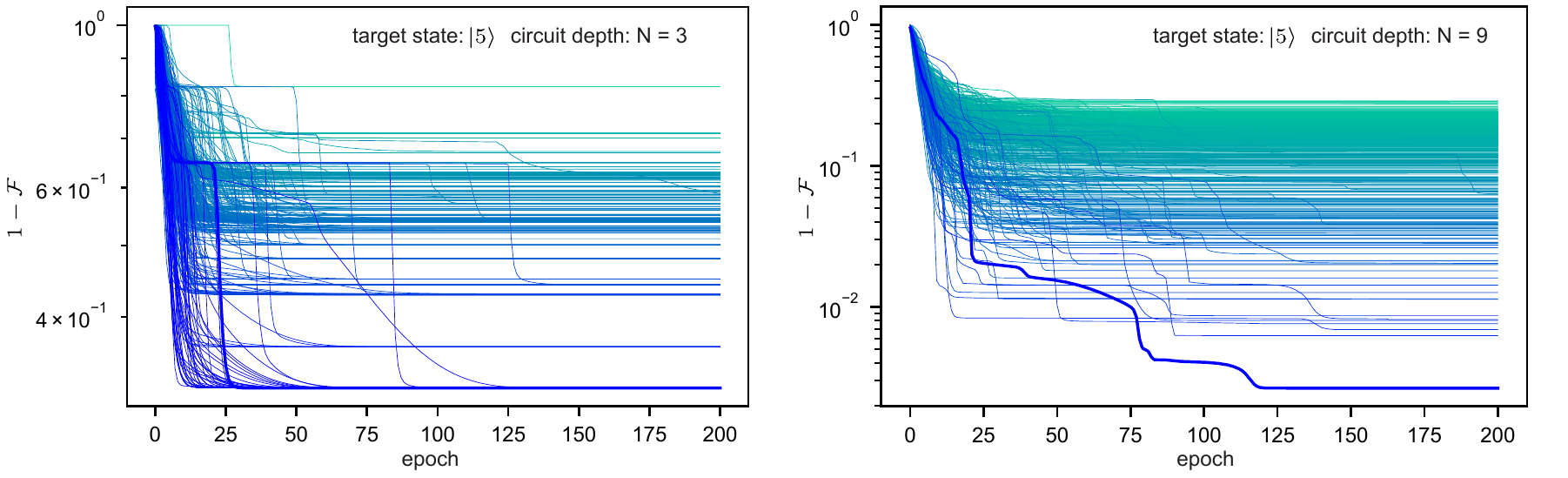}
		\caption{\label{fig:optimization} \textbf{ECD circuit optimization} Examples of optimization curves for Fock 5 state preparation. Each line represents a circuit fidelity $F_j$, and 500 randomly initialized circuits are optimized in parallel. In the case of small circuit depth $N=3$ (left panel), the best infidelity reached is $\sim 0.3$. With a larger circuit depth $N=9$ (right panel), one circuit out of the 500 reaches an  infidelity of $\sim 0.003$, demonstrating the need for multi-start optimization.}
	\end{center}
\end{figure*}

In \cref{fig:optimization} we show an example by plotting the infidelities of each circuit in the batch as a function of optimization epoch for Fock $\ket{5}$ state preparation ($\ket{0}\ket{g} \rightarrow \ket{5}\ket{g}$) using two different circuit depths, $N=3$ and $N=9$. For these examples, we use a batch size of $B=500$ circuits, with each epoch representing $100$ steps of gradient descent using Adam with a learning rate of 0.001 carried out using an Nvidia Tesla v100 GPU. 

In \cref{fig:GKP pulse}, we show an example of the pulse parameters found using this procedure for state preparation of the GKP $\ket{+Z}$ logical state. The magnitude of the echoed conditional displacements found in this example is typical of most demonstrations in this work - the largest $|\beta|$ found is $|\beta| \approx 2.75$. We note the optimization procedure does not include a constraint on $|\beta|$, but generally the scale of $|\beta|$s found is set by the phase-space extent of the target state.
As shown in this example, the optimizer often converges to pulses with interpretable values of qubit rotation angles and phases: values in that pulse are close to $\pi/4$, $\pi/2$, etc, and this is a common feature for the demonstrations in this work. 

\section{Optimization of logical gates on a finite energy GKP code}

In this section, we focus on the implementation of logical gates using ECD control. In particular, we demonstrate numerical optimization of the phase and T gates for a finite-energy square GKP encoding.
For these gates, the target state maps $\left\{\ket{\psi_{i}}\right\} \rightarrow  \left\{\ket{\psi_{t}}\right\}$ acting on the finite energy logical subspace are given by

\begin{equation}
	\label{eq:GKP state maps}
	\begin{aligned}
		&S: \{\ket{+Z}_\Delta \ket{g},\ket{-Z}_\Delta\ket{g}\} \rightarrow \{\ket{+Z}_\Delta\ket{g},e^{i\pi/2}\ket{-Z}_\Delta\ket{g}\} \\
		&T: \{\ket{+Z}_\Delta\ket{g},\ket{-Z}_\Delta\ket{g}\} \rightarrow \{\ket{+Z}_\Delta\ket{g},e^{i\pi/4}\ket{-Z}_\Delta\ket{g}\}
	\end{aligned}
\end{equation}
where we have also included the condition that the ancilla qubit starts and ends in $\ket{g}$.
To optimize these logical gates, we modify the cost function in \cref{sec:ECD optimization} to be
\begin{align}
	&C = -\sum_j \text{Re}\left(\bra{\psi_{t,j}} U_\text{ECD}\ket{\psi_{i,j}}\right)
\end{align}
where the sum is carried out over a logical state map, such as the state map for the $S$ and $T$ gates given in \cref{eq:GKP state maps}. Here, we only optimize the gate over the logical subspace, and in future work, the optimization could be modified to focus on error transparent gates \cite{maErrortransparentOperationsLogical2020}.

To quantify the quality of the optimized logical gates, we numerically calculate their average fidelity \cite{nielsenSimpleFormulaAverage2002}, defined as
\begin{align}
	\overline{\mathcal{F}} = \frac{1}{6}\text{Tr}\left(R^T[\mathcal{U}_\text{target}] R[\mathcal{E}]\right) + \frac{1}{3}
\end{align}
where $R_{ij}[\mathcal{E}] = \frac{1}{2} \text{Tr}\left(\sigma_i \mathcal{E}\left[\sigma_j\right]\right)$ is the Pauli transfer matrix (PTM) associated to a quantum channel $\mathcal{E}$. Here, we define the finite energy logical Pauli operators using the numerically computed logical states as described in \cref{sec:GKP state analysis} to be $X_\Delta = \left(\ket{+Z}_\Delta \bra{-Z}_\Delta + \ket{-Z}_\Delta \bra{+Z}_\Delta \right)\ket{g}\bra{g}$ and analogous definitions for $I_\Delta$, $Y_\Delta$, and $Z_\Delta$. For these operations, the target unitary channel is defined as $\mathcal{U}_\text{target}\left[\rho\right] = U_\text{target} \rho U^\dagger_\text{target}$ and the applied channel is $\mathcal{E}\left[\rho\right] = U_\text{ECD} \rho U^\dagger_\text{ECD}$, where $U_\text{ECD}$ is the result of the optimization.

\begin{figure*}[!ht]
	\begin{center}
		\includegraphics{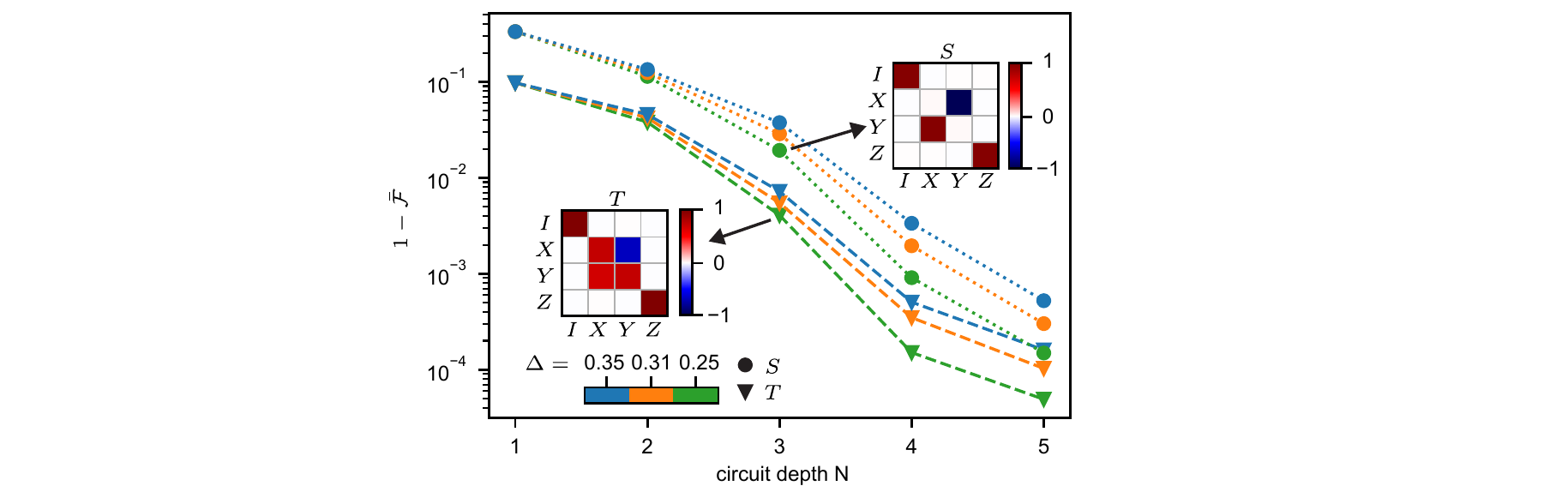}
		\caption{\label{fig:GKP gates} \textbf{Optimized finite energy S and T average fidelity.} Plotted is the average fidelity for optimized ECD circuits at different values of $\Delta$ and circuit depth $N$. S gate results are shown with circles, T gate results are indicated with triangles, and the color indicates the squeezing. Also shown (insets) are the Pauli transfer matrices for the ECD S and T gates optimized at $N=3, \Delta=0.25$. Differences in Fidelity at different values of $\Delta$ are likely caused by the small finite overlap of the finite-energy logical states.}
	\end{center}
\end{figure*}

The optimization results for the finite energy $S$ and $T$ gates are shown in \cref{fig:GKP gates} at three different squeezing values $\Delta = \left\{ 0.25, 0.31, 0.35 \right\}$. Remarkably, these gates can be performed with a low circuit depth, only requiring $N=3$ ECD gates to reach a channel fidelity $\overline{\mathcal{F}} \approx 0.99$ for the T gate and $N=4$ for the S gate. These results indicate that the ECD gate set is especially well-suited for control over the GKP code, however we note that these gate implementations are not fault tolerant with respect to ancilla qubit errors.

\section{Optimization of GRAPE and SNAP pulses}
In this section, we outline numerical methods used to generate the GRAPE and SNAP points in Fig. 2b. For these pulses, we optimize the fidelity of the quantum control problem $\ket{0}\ket{g} \rightarrow \ket{n}\ket{g}$ for the Fock states $n = 1...5$.

\subsection{Optimization of GRAPE}

We use the methods described in \cite{heeresImplementingUniversalGate2017} to optimize GRAPE pulses for the cavity and qubit using our measured $\chi$. The driven dispersive Hamiltonian is used with piecewise-constant pulses sampled every $\SI{33}{ns}$ and an oscillator Hilbert space truncation of $N = 50$. To find the numerical quantum speed limit (QSL) associated with each Fock state preparation, we sweep the length of the pulse and pick the shortest pulse with an optimized Fidelity $\mathcal{F} > 0.99$. We also employ typical bandwidth and amplitude constraints when optimizing these pulses. We note that the cavity drive amplitudes used in our ECD gates are over an order-of-magnitude larger than typical cavity drive amplitudes used in optimized GRAPE pulses \cite{heeresImplementingUniversalGate2017}. Our results confirm that GRAPE pulses optimized in the usual way take a time greater than $2\pi/\chi$ as also observed in many state-of-the art bosonic control experiments \cite{ofekExtendingLifetimeQuantum2016a,heeresImplementingUniversalGate2017,axlineOndemandQuantumState2018,huQuantumErrorCorrection2019,wangHeisenberglimitedSinglemodeQuantum2019a,wangEfficientMultiphotonSampling2020a,gertlerProtectingBosonicQubit2021,burkhartErrorDetectedStateTransfer2021, curtisSingleshotNumberresolvedDetection2021a}.

\subsection{Optimization of SNAP}
The SNAP-displacement control sequence is parameterized as $D^\dagger(\alpha) {\rm SNAP}(\varphi) D(\alpha)$ as in \cite{foselEfficientCavityControl2020}. For each target Fock state $|n\rangle$ with $n=1...5$ and circuit depth $T=1...5$, we optimize the parameters of this control sequence with reinforcement learning \cite{sivakModelFreeQuantumControl2021}. For each configuration $(n, T)$, we repeat the training 10 times with different random initial seeds. The results show that to achieve fidelity $>99\%$, the circuit depth has to be $T\ge2$ for Fock states $n=1...3$, and $T\ge3$ for $n=4...5$. The best performing protocols are compared against ECD control in the main text. For SNAP times in the main text, we assume a gate time of $2\pi /\chi$, however SNAP is typically implemented with longer gate times \cite{heeresCavityStateManipulation2015}. 

\section{Code Availability}
Code for the two-step optimization procedure - ECD parameter optimization (\cref{sec:ECD optimization}) and ECD pulse construction (\cref{sec:numerical optimization}) - is available at \url{https://github.com/alec-eickbusch/ECD_control}.

\putbib[ECD]
\end{bibunit}
\makeatother

\end{document}